\documentclass[preprint, 12pt]{elsarticle}

\usepackage[margin=3cm]{geometry}
\usepackage{lineno}
\usepackage{amsmath,amsfonts,amssymb,bm}
\usepackage[mathscr]{euscript}
\usepackage{textcomp}
\usepackage{bigints}
\usepackage[utf8]{inputenc}
\usepackage{caption}
\usepackage{subcaption}
\usepackage{multirow}
\usepackage{natbib}
\usepackage{bigints}

\newcommand{\norm}[1]{\left\lVert#1\right\rVert}
\DeclareMathOperator{\EX}{\mathbb{E}}
\modulolinenumbers[5]

\graphicspath{{Images/}}

\journal{Spatial Statistics}

\begin{document}


\begin{frontmatter}

\title{Data Fusion in a Two-stage Spatio-Temporal Model using the
  INLA-SPDE Approach}

\author{Stephen Jun Villejo\fnref{fn1,fn2}}\corref{cor1}
\ead{Stephen.Villejo@glasgow.ac.uk}

\author{Janine B Illian\fnref{fn1}}
\ead{Janine.Illian@glasgow.ac.uk}

\author{Ben Swallow\fnref{fn1}}
\ead{Ben.Swallow@glasgow.ac.uk}

\cortext[cor1]{Corresponding author}
\fntext[fn1]{School of Mathematics and Statistics, University of Glasgow, Glasgow, United Kingdom} 
\fntext[fn2]{School of Statistics, University of the Philippines Diliman, Quezon City, Philippines} 

\begin{abstract}
This paper proposes a two-stage estimation approach for a spatial misalignment scenario that is motivated by the epidemiological problem of linking pollutant exposures and health outcomes. We use the integrated nested Laplace approximation method to estimate the parameters of a two-stage spatio-temporal model – the first stage models the exposures while the second stage links the health outcomes to exposures. The first stage is based on the Bayesian melding model, which assumes a common latent field for the geostatistical monitors data and a high-resolution data such as satellite data. The second stage fits a GLMM using the spatial averages of the estimated latent field, and additional spatial and temporal random effects. Uncertainty from the first stage is accounted for by simulating repeatedly from the posterior predictive distribution of the latent field. A simulation study was carried out to assess the impact of the sparsity of the data on the monitors, number of time points, and the specification of the priors in terms of the biases, RMSEs, and coverage probabilities of the parameters and the block-level exposure estimates. The results show that the parameters are generally estimated correctly but there is difficulty in estimating the latent field parameters. The method works very well in estimating block-level exposures and the effect of exposures on the health outcomes, which is the primary parameter of interest for spatial epidemiologists and health policy makers, even with the use of non-informative priors.
\end{abstract}


\end{frontmatter}

\section{Introduction}

Spatially misaligned data, where the data are measured at different spatial scales, are a common issue in spatial modelling \cite{lawson2016handbook}. In ecological and epidemiological research for example, it is of interest to estimate the effect of pollution exposures on health outcomes such as incidence of certain diseases \cite{inlaspde, BLANGIARDO20161, LeeMukhopadhyayetal}.  Measurements of exposures are typically collected from a network of monitoring stations, from satellites, or from outputs of computer simulations using numerical models describing the origins and dispersion of exposures. Data from monitoring stations are typically sparsely located in the spatial domain, while the two latter measurements are high resolution data on fine regular grids. The former are classified as point-referenced or geostatistical data, while the latter are considered as high-resolution areal data on fine regular grids \cite{BRUNO2016276}. Moreover, data on health outcomes such as the incidence of certain diseases are available as aggregated counts on fixed and irregular areal units. Such data are organized as counts of mortality, morbidity, or hospital admissions by type of disease over a set of regions partitioning the entire study region for each time point \cite{BRUNO2016276}. Although these data derive from individual-level information, which would enable the analysis of spatial pattern of case event locations and to quantify the effects of long-term exposure, individual-level data are usually not available and are also costly to obtain in practice, and hence have been summarised across administrative units \cite{BRUNO2016276, diggle, molitor}. As a result, time series and areal study designs have been frequently used to model health outcome data. However, this comes at a cost since the summary statistics on the  population across areas cannot be used to assess effects at the individual level. Moreover, there is the danger of assuming that the associations observed from the areal level also hold at the individual level, often referred to as the \textit{ecological fallacy} \cite{BRUNO2016276}. Linking pollution exposures and health outcomes therefore necessitates inference on a spatial scale that is different from the scale of the original data, referred to as a change of support problem (COSP). 

This paper focuses on a specific spatial misalignment scenario where the outcome of interest is measured and available in areas or blocks, while the explanatory variables include both point-referenced and areal data. A sensible classical approach for this kind of spatial misalignment is to compute exposure values to represent the blocks and to then apply standard statistical methods to regress the health outcomes against the block-level exposure values \cite{BRUNO2016276}; in other words, the analysis is performed in two stages. 

In the first stage, exposure values are computed on the areas or blocks, also termed  \textit{upscaling} or \textit{smoothing}, using the points data from the monitors and the high-resolution grid data from the satellite and the dispersion models. A classical approach to the point-to-area COSP in geostatistics is block kriging \cite{gotwayoung}. However, kriging approaches require inversion of matrices, which can be computationally expensive especially for large datasets. This paper hence proposes a point-to-area upscaling method that does not require inverting large covariance matrices. If there were no high-resolution data available, a naive approach to the problem is to simply get the average of the values from the monitoring stations inside a block, possibly with the use of distance-based or population-based weights, and use this as the block-level value \cite{BRUNO2016276}. However, this approach does not work when there are blocks without point-referenced data due to the sparsity of monitoring stations in the entire spatial domain, or if the values exhibit strong spatial heterogeneity \cite{Leeetal, Kralletal}. This issue may be addressed by jointly using  satellite data and the high-resolution output of dispersion models, and data from a monitoring network; this process of merging information from different data sources is referred to as \textit{data fusion} \cite{berrocal}. One approach to performing data fusion, called \textit{Bayesian melding}, is to regard the point-referenced data and the high-resolution grid data in the first stage as outcomes from a common latent continuous random process with some measurement error, and possibly some covariates and spatial and temporal terms \cite{inlaspde, szpiropaciorek, bergenszpiro, fuentes2005model}.

In the second stage, the exposure estimates from the first stage are used as an input in a statistical model to explain the health outcomes. A naive approach that performs the first stage independently of the second stage assumes that the estimated exposure values at the block level are free from estimation error. This approach does not account for the uncertainty in the first-stage process, which can potentially lead to biased estimates of the health effects of exposures and can also underestimate the uncertainty in the second stage \cite{inlaspde,merrorspatmis}. This approach is termed the \textit{plug-in approach} and was shown to yield biased estimates, especially when the data from the monitors are sparse and the values are heterogeneous in space \cite{merrorspatmis}. An approach to overcome this issue is a two-stage Bayesian approach, in which the first stage estimates the posterior distribution of the parameters and of the latent exposures field, while the second stage estimates the health model by plugging in simulated values from the posterior predictive distribution of the first-stage model \cite{merrorspatmis, leeshaddick}. Another approach is a fully Bayesian approach which simultaneously estimates the parameters of the exposures model and the health model. However, in this framework, the health outcomes could strongly influence the exposure surface, especially if the number of monitoring stations is relatively small. Nonetheless, the two-stage Bayesian estimator is shown to approximate the fully Bayesian results quite well \cite{merrorspatmis}. 

This paper aims to propose and illustrate an approach for estimating the effect of exposures measurements, which are taken from both monitors and high-resolution grid-level data, on health outcomes measured as counts on irregular blocks. The estimation method performs data fusion and also accounts for the measurement error in the monitors and in the high-resolution proxy data using an approach derived from the Bayesian melding model. The models are latent Gaussian and are estimated using a two-stage approach using the \textit{integrated nested Laplace approximation} (INLA). Moreover, the \textit{stochastic partial differential equation} approach is used to efficiently estimate the spatial field \cite{spde, RueINLA}.  

The use of the INLA approach is motivated by its computational benefits. The INLA method is a deterministic approach for doing Bayesian inference, as opposed to \textit{Markov Chaine Monte Carlo} (MCMC) method which is a simulation-based approach \cite{RueINLA}. Rue et al. (2019) show that in particular spatial applications for which the size of the latent Gaussian field is large, the INLA method was able to compute accurate approximations of the posterior marginals which takes longer for MCMC algorithms to compute. In addition, the approximation bias is less than the MCMC error for typical examples. The SPDE approach also provides a means to speed up computation. For models using Gaussian random fields (GRF), which includes spatial and spatio-temporal models, there is a difficulty in the computation because of the need to factorize and invert typically large covariance matrices. The SPDE approach provides a Markov structure on the GRF. The Markov property causes the inverse covariance matrix, called the \textit{precision matrix}, to be sparse. This sparsity makes the factorization of the precision matrix fast. Such models are called \textit{Gaussian Markov Random Fields} (GMRF). In addition to making the computation efficient with the use of sparse matrix numerical algorithms because of the Markov structure, another benefit of using the SPDE approach is that it allows the interpretation of the model in terms of the parameters of the covariance function \cite{inlaspde}.

Section \ref{sec:methodology} reviews existing methods and ideas used in this paper. This includes a short discussion of the Bayesian melding model for data fusion, the INLA method, and the SPDE approach. Section \ref{sec:proposed.method} discusses the proposed spatio-temporal model and the assumptions of the model. It also discusses the details on how INLA and the SPDE approach can be used to fit the proposed model. Section \ref{sec:simulation.study} discusses the specifics of the simulation study, while section \ref{sec:results.discuss}
presents the results of the simulation study. 

\section{Methodology} \label{sec:methodology}

\subsection{Bayesian melding for data fusion}

Data used to conduct exposure assessment in any environmental health study come from various sources. Most studies use data from a network of monitoring stations set up and maintained by government agencies. However, since the network is typically sparse due to high maintenance costs of the stations, it will be difficult to capture spatial heterogeneity of the true exposure surface. Furthermore, networks are typically located in urban areas where the pollution level is typically high. This leads to biases when fitting models to assess the impact of exposure on health as the exposure surface will be over estimated \cite{lawson2016handbook}. 

A solution to this problem is the use of additional sources of information of exposures which provide more detailed spatial and temporal information. Two specific sources are satellite images and computer simulations using numerical or deterministic models. Satellite images are remotely sensed data and provide global coverage. On the other hand, numerical models enable the simulation of the creation and dispersion of pollution exposures using information on pollution sources and on chemical and physical processes. However, despite the rich spatial and temporal information from these data, they are prone to error. Remotely sensed data are subject to retrieval errors. Computer simulations using numerical models are sensitive to model misspecification of the underlying process, the input data on pollution sources, and the discretization of the continuous field. These two data sources are also referred to as \textit{proxy data} \cite{lawson2016handbook}. This problem of combining data from monitoring stations and proxy data from satellite images and computer simulations is referred to as \textit{data fusion} or \textit{data assimilation}. 

An approach to data assimilation, called \textit{Bayesian melding}, assumes that both the point-referenced data and the proxy data have a common latent spatial process \cite{fuentes2005model}, resulting in the following joint model:
	\begin{align}
		W(\bm{s}) &= X(\bm{s}) + e(\bm{s}) \label{eq:meldingpoint} \\
		X(\bm{s}) &= 	\mu(\bm{s}) + \epsilon(\bm{s}) \label{eq:meldinglatent} \\
		\tilde{X}(\bm{s}) &= a(\bm{s})+b(\bm{s})X(\bm{s})+\delta(\bm{s}) \label{eq:meldingproxy1}\\
		\tilde{X}(B) &= \dfrac{1}{|B|} \int_B \tilde{X}(\bm{s})d\bm{s} \label{eq:meldingproxy2}
	\end{align}
$W(\bm{s})$ denotes the point-referenced random outcome at location $\bm{s}$, $\tilde{X}(B)$ the outcome of the proxy data in grid $B$, and $X(\bm{s})$ the latent spatial stochastic process. Equation \eqref{eq:meldingpoint} states that the observed values at the monitoring stations are error-prone realizations of the latent process with $e(\textbf{s})$ as the random error term. Equation \eqref{eq:meldinglatent} postulates that the latent spatial stochastic process is decomposed into a mean process $\mu(\bm{s})$ and a residual process $\epsilon(\bm{s})$. The mean of $X(\bm{s})$ can be a function of covariates, while the residual component may include spatial effects. Equation \eqref{eq:meldingproxy1} models the proxy data at the point-level. The point-level outcome $\tilde{X}(\bm{s})$ is associated with the latent process as a linear function of $X(\bm{s})$, with $a(\bm{s})$ and $b(\bm{s})$ as additive and multiplicative calibration parameters, respectively. In addition to that, there is also an additional random error term, $\delta(\bm{s})$, for $\tilde{X}(\bm{s})$. The calibration parameters $a(\bm{s})$ and $b(\bm{s})$ can be assumed to be constant or varying over space, but they are usually considered as fixed effects to avoid issues with identifiability when estimating $X(\bm{s})$. Finally, equation \eqref{eq:meldingproxy2} defines the observed value of $\tilde{X}(B)$ as a spatial average of $\tilde{X}(\bm{s})$. Studies have shown that Bayesian melding outperforms kriging in predicting pollution fields \cite{berrocal2010spatio, liu2011empirical}. The idea of assuming a common latent spatial process for the observed data at different spatial scales was also used in \cite{wikle2005combining, mcmillan2010combining, sahu2010fusing}. 

\subsection{INLA for latent Gaussian models}

Rue et al.\ (2009) proposed a method for approximate Bayesian inference for latent Gaussian models called integrated nested Laplace approximation (INLA) \cite{RueINLA}. Latent Gaussian models are a subset of the class of Bayesian additive models with a structured additive predictor. Suppose $y_i$ is the response variable with distribution function in the exponential family, and with mean $\mu_i$ which is linked to an additive predictor $\eta_i$ through a link function $g(.)$, i.e.\ $g(\mu_i)=\eta_i$, and possibly other parameters say $\bm{\theta}_2$. The general form of $\eta_i$ is 
\begin{equation}\label{eq:lgm}
	\eta_i = \alpha + \sum_{j=1}^{n_f}f^{(j)}(u_{ij}) + \sum_{k=1}^{n_{\beta}}\beta_k z_{ik} + \epsilon_i.
\end{equation}
In \eqref{eq:lgm}, $\alpha$ is the intercept, $\{\beta_k\}$'s are the coefficients of known covariates $\bm{z}$, $\epsilon_i$ is an error term, and $\{f^{(j)}(.)\}$'s are unknown functions of covariates $\bm{u}$. The covariates $\bm{u}$ can be spatial points or time points so that the $f^{(j)}(u_{ij})$ terms take care of the spatial, temporal, or spatio-temporal dynamics in the model. The functions $f^{(j)}$ can take several forms to relax the linear relationship with the covariate. It can be modelled using parametric nonlinear terms, nonparametric forms such as a random walk model, or Gaussian processes. All unknown quantities in equation \eqref{eq:lgm}, namely $\alpha, \{\beta_k\}, \{f^{(j)}(\cdot)\}, \{\epsilon_i\}$ are assumed to be Gaussian, and thus the model is called a \textit{latent Gaussian model}. The elements of the latent Gaussian space, put into a vector $\bm{x}$, depends on another set of parameters $\bm{\theta}_1$.

The INLA method aims to estimate the marginal posterior distributions of each element of $\bm{x}$ and $\bm{\theta} = \begin{pmatrix}
\bm{\theta}_1^{\intercal} & \bm{\theta}_2^{\intercal}
\end{pmatrix}^{\intercal}$, which are derived as follows:
\begin{equation}\label{latentmarg}
\pi(x_i|\bm{y}) = \int \pi(x_i|\bm{\theta},\bm{y})\pi(\bm{\theta}|\bm{y}) d\bm{\theta}
\end{equation}
\begin{equation} \label{thetamarg}
\pi(\theta_i|\bm{y}) = \int \pi(\bm{\theta}|\bm{y})d\bm{\theta}_{-j}
\end{equation}

To estimate the posterior marginals, $\pi(x_i|\bm{\theta},\bm{y})$ and $\pi(\bm{\theta}|\bm{y})$ are first approximated using Laplace approximations. Given the respective approximations, $\tilde{\pi}(x_i|\bm{\theta},\bm{y})$ and $\tilde{\pi}(\bm{\theta}|\bm{y})$, equation \eqref{latentmarg} is approximated using numerical integration via
\begin{equation}\label{numint}
	\tilde{\pi}(x_i|\bm{y}) = \sum_{k}\tilde{\pi}(x_i|\bm{\theta}^*_k,\bm{y})\tilde{\pi}(\bm{\theta}^*_k|\bm{y})\Delta_k,
\end{equation}

where $\{\bm{\theta}^*_k\}$ are evaluation points and $\{\bm{\Delta}_k \}$ are the corresponding weights. Performing the numerical integration in equation \eqref{numint} requires good evaluation points. Identifying the evaluation points can be done by a grid search. However, this can be computationally expensive even for a moderately large dimension of $\bm{\theta}$. An alternative approach is to frame the problem as a design problem which then gives fewer integration points than the grid search. The naive approach, called the \textit{empirical Bayes} approach, is to simply plug-in the mode of $\tilde{\pi}(\bm{\theta}|\bm{y})$ in equation \eqref{numint}. Although this is the fastest approach, this ignores the uncertainty in the $\bm{\theta}$-space and thus underestimates the overall variability. 

\subsection{SPDE approach} \label{subsec:spde}

Lindgren and Rue (2011) showed a link between Gaussian fields (GF) with covariance function of the Mat\`{e}rn type with a GMRF whose precision matrix approximates well the inverse covariance matrix of the GF in some norm using a stochastic partial differential equation (SPDE) \cite{spde}. The Mat\`{e}rn covariance function between locations $\bm{s}_i$ and $\bm{s}_j$ in the $d$-dimensional space is given by
	\begin{equation} \label{eq:maternfunction}
		C(\bm{s}_i,\bm{s}_j)=\dfrac{\sigma^2}{2^{\nu-1}\Gamma(\nu)} (\kappa\norm{\bm{s}_i-\bm{s}_j})^{\nu}K_{\nu}(\kappa(\norm{\bm{s}_i-\bm{s}_j}))
	\end{equation}
	where $\norm{\cdot}$ is the Euclidean distance in $\mathbb{R}^d$. In the Mat\`{e}rn equation, $\sigma^2$ is the marginal variance, $K_{\nu}$ is the modified Bessel function of the second kind and order $\nu>0$, and $\kappa>0$ is a scaling parameter. The parameter $\nu$ defines the smoothness of the underlying process and is usually fixed at some value. The scaling parameter, $\kappa$, is related to the range parameter $\rho$, which is the distance at which the correlation is around 0.1. The empirically derived relationship is given by $\rho=\dfrac{\sqrt{8\nu}}{\kappa}$.

The stochastic partial differential equation is given by
	\begin{equation}\label{spde}
		(\kappa^2-\Delta)^{\alpha/2}x(\bm{s}) = \mathscr{W}(\bm{s}), \;\;\;\; \bm{s}\in\mathbb{R}^d, \;\;\; \alpha = \nu+d/2, \;\;\; \kappa>0, \;\;\; \nu>0
	\end{equation}
	where $(\kappa^2-\Delta)^{\alpha/2}$ is a pseudodifferential operator, $\mathscr{W}(\bm{s})$ is a Gaussian white noise with unit variance, and $\Delta$ is the \textit{Laplacian} defined by $\Delta = \sum_{i=1}^{d} \dfrac{\partial^2}{\partial x_i^2}$.

	The solution of the spde in equation \eqref{spde}, which is a GF with Mat\`{e}rn covariance function, can be approximated by a Mat\`{e}rn Gaussian random field. The approximation uses a discretization process called the \textit{Finite Element Method} which starts by defining a triangulation in $\mathbb{R}^d$ \cite{spde, ciarlet2002finite}. The observed locations are usually used as the initial  \textit{vertices} in the triangulation, and then additional vertices are added given some constraints such as maximum edge length and minimum interior angles. The next step is to define a basis function representation of $x(\bm{s})$ using the triangulation in $\mathbb{R}^d$, given by
	\begin{equation}\label{basisrep}
		x(\bm{s}) = \sum_{k=1}^{K}\psi_k(\bm{s})w_k
	\end{equation}
	for some basis functions $\{\psi_k\}$ and Gaussian-distributed weights $\{w_k\}$, and where $K$ is the number of vertices. The basis functions are chosen to be piecewise linear in each triangle, and so $\psi_k=1$ at vertex $k$ and 0 at other vertices. The joint distribution of the weights $w_k$ determines the full distribution of the continuously indexed but finite-dimensional solution given by equation \eqref{basisrep}. 

The Mat\`{e}rn parameter $\nu$ is usually fixed at some value with a default value of 1. In the two-dimensional space ($d=2$), this corresponds to $\alpha = 2$. When $\alpha = 2$, the precision matrix of the Gaussian weights $\{w_k\}$ is given by
	\begin{equation}\label{computeQ}
	\bm{Q}=\kappa^4\bm{C}+2\kappa^2\bm{G}+\bm{G}\bm{C}^{-1}\bm{G}
	\end{equation}
	where $C_{ij} = \langle \psi_i, \psi_j \rangle$ and $G_{ij}=\langle \nabla\psi_i, \nabla\psi_j \rangle$, $\nabla$ being the gradient. Hence, for a given $\kappa$ and $\alpha$ value in the SPDE, we have a mapping from the parameters of the GF Mat\`{e}rn model to the elements of the precision matrix $\bm{Q}$. The link between the SPDE parameters and the Mat\`{e}rn covariance function parameters are discussed in \cite{rinlabook}. $\bm{C}$ and $\bm{G}$ are sparse matrices since it has non-zero elements only for pairs of $\psi_i$ and $\psi_j$ which share common triangles; hence, the precision matrix $\textbf{Q}$ is sparse as well. Hence, the approximate solution to the SPDE is a GMRF, where the Gaussian weights $w_1,w_2,...,w_K$ are multivariate Gaussian with mean $\mathbf{0}$ and sparse precision matrix $\bm{Q}$. The elements of $\bm{Q}$ for other positive integer values of $\alpha$ are discussed in \cite{RueMart}. Some asymptotic results of the approximate GMRF representation of the Mat\`{e}rn field can also be found in \cite{RueMart}.

\section{Proposed method} \label{sec:proposed.method}

\subsection{Model assumptions} \label{subsec:modelassumptions}

There are two main model structures, one for the exposures (referred to as the first-stage model) and another one to link the health outcomes with the exposures (the second-stage model). As will be discussed later, the model for the exposures is defined continuously in space. An important model assumption from this is that spatial averages of this continuous stochastic process, computed on blocks where the counts of health outcomes are observed, are inputs in the second-stage model. Hence, in the proposed method as discussed in sections \ref{subsec:proposed.stage1.model} and \ref{subsec:proposed.stage2.model}, the estimated first-stage model is used to interpolate the values of the latent process over the entire study region. These values are then used to compute block-level estimates of the exposures using spatial averaging. The spatial averages are used as predictors in the second-stage model for the observed health outcomes. When estimating the effect of exposures on the health outcomes using the two-stage modelling framework, it is important to account for the estimation error in the block-level estimates of exposures. This issue is discussed in section \ref{subsec:propagate.uncertainty}. 

\subsubsection{First-stage model}

The first-stage model follows the spatio-temporal model proposed in \cite{cameletti2013spatio} for particulate matter concentration in the North-Italian region Piemonte.  Suppose $x(\bm{s},t)$ is the true exposures value at location $\bm{s}$ and time $t$ and $\bm{z}(\bm{s},t)$ are the covariates. The model for the exposures surface is given by 
\begin{align}
				x(\bm{s},t) &= \beta_0 + \bm{\beta}^{\intercal}\bm{z}(\bm{s},t)+\xi(\bm{s},t) \\
				&\xi(\textbf{s},t) = \varsigma \xi(\bm{s},t-1) + \omega(\bm{s},t), \;\;\; |\varsigma| < 1, t=2,...,T,
\end{align}
where $\beta_0$ is the intercept of the model, and $\bm{\beta}$ is the coefficient vector of the covariates $\bm{z}(\bm{s},t)$. The spatio-temporal dependence in the model is induced by $\xi(\bm{s},t)$ which evolves in time as an autoregressive process of order 1, with $\varsigma$ as the autoregressive parameter and $\xi(\bm{s},1) \sim N(0,\sigma^2_{\omega}/(1-\varsigma^2))$. The term $\omega(\bm{s},t)$ is a temporally-independent Gaussian random field  with mean 0 and Mat\`{e}rn covariance function, that is
\begin{align}
				\text{Cov}(\omega(\bm{s}_i,t),\omega(\bm{s}_j,u)) = \begin{cases} 
  						    0 & t \neq u \\
  						    \Sigma_{i,j}& t = u
  							 \end{cases} 
\end{align}
where $\Sigma_{i,j}$ takes the form in equation \eqref{eq:maternfunction} with $\nu = 1$. The parameters of the Mat\`{e}rn covariance function are $\sigma^2_{\omega}$ and $\kappa$ which refers to the spatial variance and the scaling parameter, respectively. The interpretation of the spatio-temporal structure of the model is usually in terms of the $\sigma^2_{\omega}$ and the range parameter $\rho$. The empirically derived relationship between $\rho$ and $\kappa$ is discussed in section \ref{subsec:spde}. 

Following the classical error model, the observed values at the monitors are error-prone realizations of the true exposures surface. This also follows equation \eqref{eq:meldingpoint} of the Bayesian melding model. Suppose there are $M$ monitors and the observed value at a monitor in location $\bm{s}_i$ at time $t$ is denoted by $w(\bm{s}_i,t)$. The model for the observed values at the monitors is
				\begin{align}
					w(\bm{s}_i,t) &= x(\bm{s}_i,t)+e(\bm{s}_i,t), \;\;\; e(\bm{s},t) \sim N(0,\sigma^2_e), \;\;\;i=1,\ldots,M 
				\end{align}
The model for the proxy data in the Bayesian melding model is defined in equations \eqref{eq:meldingproxy1} and \eqref{eq:meldingproxy2}, so that the observed outcome at a grid $B$ at time $t$, denoted by $\tilde{x}(B,t)$, would be 
            \begin{align} \tilde{x}(B,t) &= \dfrac{1}{|B|}\int_{\forall \bm{s} \in B} \tilde{x}(\bm{s},t) \\ &=\dfrac{1}{|B|}\int_{\forall \bm{s} \in B}  \bigg( \alpha_0 + \alpha_1  x(\bm{s},t)+ \delta(\bm{s},t) \bigg) d\bm{s}, \;\;\; \delta(\bm{s},t) \sim N(0,\sigma_{\delta}^2)
			  \end{align}
which is simulated as 
		  	\begin{align} \tilde{x}(B,t) &=  \dfrac{1}{\#(\bm{s}\in B)}\sum_{\forall \bm{s} \in B}  \bigg( \alpha_0 + \alpha_1  x(\bm{s},t)+ \delta(\bm{s},t) \bigg),
			  \end{align} 
where $|B|$ is the size of the grid and $\#(\bm{s}\in B)$ is the number of simulated points inside the grid. However, a full Bayesian melding model is not adopted in this work. With the assumption that the resolution of the proxy data is very fine, the proxy data is also treated as geostatistical at the centroids. This means that 
\begin{align} \label{eq:simplification}
    \tilde{x}(B,t) = \tilde{x}(\bm{s},t) = \alpha_0 + \alpha_1  x(\bm{s},t)+ \delta(\bm{s},t), \;\;\; \bm{s} \text{ is the centroid of } B
\end{align}
Nonetheless, just like the full Bayesian melding model, the proposed simplification still incorporates an additive bias given by $\alpha_0$, a multiplicative bias given by $\alpha_1$, and an additional zero-mean random noise term with variance $\sigma_{\delta}^2$ in the proxy data. This accounts for the fact that the proxy data have more noise and are less correlated with the true latent field as compared with the observed values from the monitors \cite{lawson2016handbook}. The proposed method is, therefore, not a full Bayesian melding approach since both the proxy data and the observed data from the monitors are considered as geostatistical data, which we conveniently call a \textit{non fully Bayesian melding model}. A particular scenario for a full Bayesian melding approach to be infeasible is when there is no available covariate information whose resolution is finer than the resolution of the proxy data. For instance, if the resolution of the proxy data is the same for the covariate information $\bm{z}(\bm{s},t)$, then it will not be feasible to estimate equations \eqref{eq:meldingproxy1} and \eqref{eq:meldingproxy2} unless it will be assumed that the value of $\bm{z}(\bm{s},t)$ for any point $\bm{s}$ inside $B$ is constant. Thus, an implicit assumption in the data generating process is that the covariate information on $\bm{z}(\bm{s},t)$ is collected at the same resolution as the proxy data and also at the level of the network of monitors. This implies that the resolution of the prediction grid is also of the same resolution as the proxy data, since any predictions from the estimated latent field need the covariate information. Also, as would be presented in the details of the simulation study, the resolution of the simulation grid for the latent field is the same as the resolution of the proxy data but incorporating biases from different sources. This simplification in the model assumptions, via the use of a \textit{non-fully Bayesian melding model}, also simplifies the estimation procedure as would be presented in section \ref{subsec:proposed.stage1.model}. The possible issues with the use of a full Bayesian melding approach in as far as estimation is concerned via the INLA method is discussed in the succeeding sections.

Lastly, the true value of the exposures at an area or block $B_i$ at time $t$, denoted by $x(B_i,t)$, is a spatial  average of the process $x(\bm{s},t)$ for all $x(\bm{s},t) \in B_i$, i.e., 
        \begin{align} \label{eq:spatialaverage}
					x(B_i,t) = \dfrac{1}{|B_i|}\int_{B_i} x(\bm{s},t)d\textbf{s}.
				\end{align}
In terms of the notation, $x(B_i,t)$ denotes a spatial average of the latent process, while $\tilde{x}(B,t)$ is the observed value of the proxy at a grid $B$ which is assumed to be equal to $\tilde{x}(\bm{s},t)$, where $\bm{s}$ is the centroid of $B$, as shown in equation \eqref{eq:simplification}.

\subsubsection{Second-stage model}

The second-stage model specifies the model for the health outcomes as the response and the block-level spatial averages of $x(\bm{s},t)$ defined in equation \eqref{eq:spatialaverage} as an input. The observed count at block $B_i$ at time $t$, denoted by $Y(B_i,t)$, is assumed to be a Poisson outcome, i.e.,
				\begin{align}
					Y(B_i,t) \sim \text{Poisson}\big(P(B_i,t)\lambda(B_i,t)\big),
				\end{align}
				where $P(B_i,t)$ is the number of expected cases in area $B_i$ at time $t$, and $\lambda(B_i,t)$ is the relative risk modelled as
				\begin{align}
				\log(\lambda(B_i,t)) = \gamma_0 + \gamma_1 x_1(B_i,t) + \varphi_{it},
				\end{align}
where $\gamma_0$ is the intercept, $\gamma_1$ is the coefficient of the true block-level exposure $x(B_i,t)$, and $\varphi_{it}$ is a spatio-temporal random effect. The spatio-temporal random effect term can take several forms discussed in \cite{rinlabook}. A general form is given as
				\begin{align}
				\log(\lambda(B_i,t)) = \gamma_0 + \gamma_1 x(B_i,t) + \phi_i + \psi_i + \upsilon_t + \nu_t + \Phi_{it},
				\end{align}
where $\phi_i$ is an \textit{iid} spatial random effect, $\psi_i$ is a structured spatial random effect, $\upsilon_t$ is an \textit{iid} temporal random effect, $\nu_t$ is a structured spatial random effect, and $\Phi_{it}$ is a spatio-temporal interaction effect. The interaction effect has four types, depending on which of the two spatial effects and which of the two time effects interact. Structured random effects include random walk models, autoregressive processes, and correlated random effects of different dimensions. These different ways to specify the spatio-temporal structure of the model do not violate latent Gaussianity, and hence all posterior marginals of the model parameters can be fit using the INLA method \cite{rinlabook}. 

For the purpose of illustrating how INLA can be used to fit a spatio-temporal model of such form, this paper assumes a simple model structure for the spatial and temporal random effects. In the simulation study done in this paper, it is assumed that 	                           \begin{align}
				\log(\lambda(B_i,t)) = \gamma_0 + \gamma_1 x(B_i,t) + \phi_i + \nu_t,
				\end{align} 
where $\phi_i$ is an $iid$ random effect and $\nu_t$ follows a random walk model of order 1, i.e., $\phi_i \sim \text{N}(0,\sigma_{\phi}^2)$ and $ \nu_t|\nu_{t-1} \sim N(\nu_{t-1},\sigma^2_{\nu})$. Sections \ref{subsec:proposed.stage1.model} and \ref{subsec:proposed.stage2.model} discuss the details of how the models are fit using INLA. All the ideas and the methods also apply for more complex spatio-temporal effects as long as the latent Gaussianity of the model is not violated.

\subsection{Model fitting for first-stage model} \label{subsec:proposed.stage1.model}

\subsubsection{Derivation of posteriors}

Suppose we arrange the data from the monitors as $\bm{w} = \begin{pmatrix} \bm{w}_1^\intercal & \bm{w}_2^\intercal & \ldots & \bm{w}_T^\intercal \end{pmatrix}^\intercal$, where $\bm{w}_t^\intercal = \begin{pmatrix} w(\bm{s}_1,t) & w(\bm{s}_2,t)& \ldots & w(\bm{s}_M,t) \end{pmatrix}, t= 1,\ldots,T.$ We assume that we have $T$ time points and $M$ monitors. Also, the data from the single proxy data is given by $\tilde{\bm{x}} = \begin{pmatrix} \tilde{\bm{x}}_1^\intercal & \tilde{\bm{x}}_2^\intercal & \ldots & \tilde{\bm{x}}_T^\intercal \end{pmatrix}^\intercal$, where $\tilde{\bm{x}}_t^\intercal = \begin{pmatrix} \tilde{x}(\bm{g}_1,t) & \tilde{x}(\bm{g}_2,t)& \ldots & \tilde{x}(\bm{g}_G,t) \end{pmatrix}$, $ t= 1,\ldots,T.$ Here, we denote by $G$ the number of grids for the proxy data, and we denote by $\tilde{x}(\bm{g}_i,t)$ the observed value at the grid whose centroid is $\bm{g}_i$ at time $t, i=1,\ldots,G$. Since both $\bm{w}_t$ and $\tilde{\bm{x}}_t$ are error-prone realizations of the true exposure values $\bm{x}_t$, then we define the vector of true exposures combined for both the monitors and the grids of the proxy data at time $t$ as
\begin{align}
 \bm{x}_t = \begin{pmatrix} \bm{x}_{t,M} & \bm{x}_{t,P}\end{pmatrix}^\intercal,
\end{align}
where $\bm{x}_{t,M}$ denotes the vector of true exposures at the monitors at time $t$, and $ \bm{x}_{t,P}$ denotes the vector of true exposures at the centroids of the grids of the proxy data at time $t, t=1,\ldots,T$. Then the vector of true exposures for all $t=1,\ldots,T$ is denoted by
$\bm{x} = \begin{pmatrix} \bm{x}_1^\intercal & \bm{x}_2^\intercal & \ldots & \bm{x}_T^\intercal \end{pmatrix}^\intercal$. Similarly, we define $\bm{\xi} = \begin{pmatrix} \bm{\xi}_1^\intercal & \bm{\xi}_2^\intercal & \ldots & \bm{\xi}_T^\intercal \end{pmatrix}^\intercal$, where $\bm{\xi}_t^\intercal = \begin{pmatrix} \bm{\xi}_{t,M}^\intercal & \bm{\xi}_{t,P}^\intercal \end{pmatrix}$, with $\bm{\xi}_{t,M}$ as the vector of spatio-temporal random effects at the monitors for time $t$ and $\bm{\xi}_{t,P}$ as the vector of spatio-temporal random effects at the centroids of the grids of the proxy data at time $t$. And finally $\bm{\omega} = \begin{pmatrix} \bm{\omega}_1^\intercal & \bm{\omega}_2^\intercal & \ldots & \bm{\omega}_T^\intercal \end{pmatrix}^\intercal$ as the vector of values of the Gaussian random field, where $\bm{\omega}_t^\intercal =  \begin{pmatrix} \bm{\omega}_{t,M}^\intercal & \bm{\omega}_{t,P}^\intercal \end{pmatrix}$; and $\bm{z} = \begin{pmatrix} \bm{z}_1^\intercal & \bm{z}_2^\intercal & \ldots & \bm{z}_T^\intercal \end{pmatrix}^\intercal$, as the vector of a single covariate, where $\bm{z}_t^\intercal =  \begin{pmatrix} \bm{z}_{t,M}^\intercal & \bm{z}_{t,P}^\intercal \end{pmatrix}$. This is can easily be generalized to the case of more than one covariate. 

The first-stage estimation procedure fits the following joint model:
\begin{align}
	\bm{w}_t &= \bm{x}_{t,M} + \bm{e}_t, \;\;\; \bm{e}_t\sim N(\textbf{0},\sigma^2_{e}\mathbb{I}_M), \;\; t=1,\ldots, T \label{eq1}\\
	\tilde{\bm{x}}_t &= \alpha_0\bm{1}_G + \alpha_1\bm{x}_{t,P} + \bm{\delta}_t,\;\;\;\bm{\delta}_t\sim N(\mathbf{0},\sigma^2_{\delta} \mathbb{I}_G), \;\; t=1,\ldots, T \label{eq2}\\
	\begin{pmatrix} \bm{x}_{t,M} \\ \bm{x}_{t,P} \end{pmatrix}&= \beta_{0}\mathbf{1}_{M+G} + \beta_{1} \begin{pmatrix} \bm{z}_{t,M} \\ \bm{z}_{t,P} \end{pmatrix} + \begin{pmatrix} \bm{\xi}_{t,M} \\ \bm{\xi}_{t,P} \end{pmatrix}, \;\; t=1,\ldots, T \label{eq3} \\				
	\begin{pmatrix} \bm{\xi}_{t,M} \\ \bm{\xi}_{t,P} \end{pmatrix} &= \varsigma \begin{pmatrix} \bm{\xi}_{t-1,M} \\ \bm{\xi}_{t-1,P} \end{pmatrix} + \begin{pmatrix} \bm{\omega}_{t,M} \\ \bm{\omega}_{t,P} \end{pmatrix}, \;\begin{pmatrix} \bm{\omega}_{t,M} \\ \bm{\omega}_{t,P} \end{pmatrix}\sim N(\mathbf{0}, \bm{\Sigma}),\;\; t=1,\ldots, T
\end{align}		
where $\mathbb{I}_M$ and $\mathbb{I}_G$ are identity matrices of dimension $M$ and $G$, respectively, and $\bm{1}_G$ and $\bm{1}_{M+G}$ is a vector of 1's of dimension $G$ and $M+G$, respectively. As discussed in section \ref{subsec:modelassumptions}, $\bm{\omega}_t$ is a temporally-independent Gaussian vector with mean zero and covariance matrix $\bm{\Sigma}$, whose elements are computed using the Mat\`{e}rn covariance function with parameters $\sigma^2_{\omega}$ and $\kappa$. 

In the system of equations above, the latent vector $\bm{x}_{t,M}$ is present in both \eqref{eq1} and \eqref{eq3}. Also, $\bm{x}_{t,P}$ is present in both \eqref{eq2} and \eqref{eq3}. In order to make sure that the values of $\bm{x}_{t,M}$ and $\bm{x}_{t,P}$ are equivalent for the different equations when fitting the joint model, $\bm{x}_{t,M}$ in \eqref{eq1} is assumed to be an (almost) identical copy of $\bm{x}_{t,M}$ in \eqref{eq3}. Similarly, $\bm{x}_{t,P}$ in \eqref{eq2} is assumed to be an (almost) identical `copy' of $\bm{x}_{t,P}$ in \eqref{eq3} where $\alpha_1$ is a scaling parameter. To create these `copies', the latent field $\bm{x}_t$ is extended to $\bm{\chi}_t = \begin{pmatrix} \bm{x}_t^{\intercal} & \bm{x}^{*\intercal}_t \end{pmatrix}^\intercal, t=1,\ldots,T$, where $\bm{x}^*_t = \begin{pmatrix} \bm{x}^{*\intercal}_{t,M} & \bm{x}^{*\intercal}_{t,P} \end{pmatrix}^\intercal$ is a copy of $\bm{x}_t$. The prior specification for the extended latent field at time $t$, $\pi(\bm{\chi}_t)$, will ensure that $\bm{x}^*_t$ is an identical copy of $\bm{x}_t$. In particular, $\bm{x}^{*}_{t,M}$ and $\bm{x}^{*}_{t,P}$ will be defined later in such a way that $\EX(\bm{x}^{*}_{t,M}) = \bm{x}_{t,M}$ and $\EX(\bm{x}^{*}_{t,P}) = \alpha_1 \bm{x}_{t,P}$, where $\alpha_1$ is a scaling parameter. This is the same approach as in \cite{Martetal} and \cite{ruiz2012direct} and is called a \textit{data augmentation approach}.

Since $\bm{x}_t$ on the left-hand side of \eqref{eq3} is unknown, $\bm{x}_t$ is transposed on the right-hand side of the equation. The remaining vector of zeroes in the left-hand side are referred to as `pseudo-zeroes'. This gives the following re-expression of the joint model:
\begin{align}
	\bm{w}_t &= \bm{x}^*_{t,M} + \bm{e}_t, \;\;\; \bm{e}_t\sim N(\textbf{0},\sigma^2_{e}\mathbb{I}_M), \;\; t=1,\ldots, T \\\
	\tilde{\bm{x}}_t &= \alpha_0\bm{1}_G + \bm{x}^{*}_{t,P} + \bm{\delta}_t,\;\;\;\bm{\delta}_t\sim N(\mathbf{0},\sigma^2_{\delta} \mathbb{I}_G), \;\; t=1,\ldots, T \\
\bm{0}_t &= -\begin{pmatrix} \bm{x}_{t,M} \\ \bm{x}_{t,P} \end{pmatrix}+ \beta_{0}\mathbf{1}_{M+G} + \beta_{1}\bm{z}_t+ \bm{\xi}_t,\;\; t=1,\ldots, T \label{pseudozeroes} \\				\bm{\xi}_t &= \varsigma \bm{\xi}_{t-1} + \bm{\omega}_t, \;\;\; \bm{\omega}_t \sim N(\mathbf{0}, \bm{\Sigma}),\;\; t=1,\ldots, T
\end{align}	

Suppose we have $\bm{\theta} = \begin{pmatrix} \sigma^2_e & \sigma^2_{\delta} &\alpha_0 & \alpha_1 & \beta_0 & \beta_1 & \varsigma & \sigma^2_{\omega} & \kappa \end{pmatrix}^\intercal$. The posterior distribution of interest is $\pi(\bm{\chi}, \bm{\xi},\bm{\theta}|\bm{w}, \tilde{\bm{x}},\bm{0})$, given by 
\begin{align}
	\pi(\bm{\chi}, \bm{\xi},\bm{\theta}|\bm{w}, \tilde{\bm{x}},\bm{0}) \propto \pi(\bm{w}|\bm{\chi},\bm{\xi},\bm{\theta})  \pi(\tilde{\bm{x}}|\bm{\chi},\bm{\xi},\bm{\theta}) \pi(\bm{0}|\bm{\chi},\bm{\xi},\bm{\theta}) \pi(\bm{\xi}|\bm{\theta}) \pi(\bm{\theta}) \pi(\bm{\chi}).
\end{align}

The form of each component is given in \ref{app:stage1posteriors}. The first two is straightforward since $\bm{w}_t|\bm{\chi},\bm{\xi},\bm{\theta} \overset{\mathrm{iid}}{\sim} N(\bm{x}^*_{t,M}, \sigma^2_{e}\mathbb{I}_M).$ Also, $\tilde{\bm{x}}_t|\bm{\chi},\bm{\xi},\bm{\theta} \overset{\mathrm{iid}}{\sim} N(\alpha_0\bm{1}_G + \bm{x}^*_{t,P}, \sigma^2_{\delta} \mathbb{I}_G).$

For the pseudo-zeroes, we have 
\begin{align}
\mathbf{0}_t|\bm{\chi},\bm{\xi},\bm{\theta} \sim N(-\bm{x}_t + \beta_{0}\bm{1}_{M+G} + \beta_{1}\bm{z}_t+ \bm{\xi}_t, \tfrac{1}{\tau_0}\mathbb{I}_G),
\end{align} 
where $\tau_0$ is a precision parameter and is fixed at a large value because of the absence of measurement error in the pseudo-zeroes.

The form of the distribution of $\bm{\xi_t}|\bm{\theta}$ uses the fact that $\bm{\xi}_t|\bm{\xi}_{t-1}\sim N(\varsigma \bm{\xi}_{t-1},\bm{\Sigma}), t=2,\ldots,T$, and that $\bm{\xi}_1 \sim N(\mathbf{0}, \tfrac{1}{1-\varsigma^2}\bm{\Sigma})$. 

For the extended latent field $\bm{\chi}$, we have
\begin{align}
	\pi(\bm{\chi}) &= \prod_{t=1}^T \pi(\bm{\chi}_t) \\
	&=\prod_{t=1}^T  \pi(\bm{x}_t^*|\bm{x}_t) \pi(\bm{x}_t) \\
	&=\prod_{t=1}^T  \pi(\bm{x}_{t,M}^*|\bm{x}_t) \pi(\bm{x}_{t,P}^*|\bm{x}_t) \pi(\bm{x}_t)\\
	&=\prod_{t=1}^T  \pi(\bm{x}_{t,M}^*|\bm{x}_{t,M}) \pi(\bm{x}_{t,P}^*|\bm{x}_{t,P}) \pi(\bm{x}_t)
\end{align}

Since $\bm{x}_{t,M}^*$ and $\bm{x}_{t,P}^*$ are independent copies of $\bm{x}_{t,M}$ and $\bm{x}_{t,P}$, respectively, then both are assumed to be Gaussian centered on $\bm{x}_{t,M}$ and $\alpha_1\bm{x}_{t,P}$ and with very high precision, i.e., $\bm{x}_{t,M}^*|\bm{x}_{t,M}\sim N\bigg(\bm{x}_{t,M}, \dfrac{1}{\tau_{x^*}}\bigg)$ and $\bm{x}_{t,P}^*|\bm{x}_{t,P}\sim N\bigg(\alpha_1 \bm{x}_{t,P}, \dfrac{1}{\tau_{x^*}}\bigg)$,  where $\tau_{x^*}$ is fixed at some large value. The latent field $\bm{x}_t$ is then assumed to be independent Gaussian centered at zero but with fixed high value for variance (low precision), i.e., $\bm{x}_t \sim N\bigg(\mathbf{0}, \dfrac{1}{\tau}_x\bigg)$, where $\tau_x$ is a small value. Although the precision is very low, the pseudo-zeroes have very high precision, and so the value of $\bm{x}_t$ in \eqref{pseudozeroes} is forced to be close to its true value. 
 
 Finally, the components of $\bm{\theta}$ can be safely assumed to be independent, i.e., $\pi(\bm{\theta})=\prod_{i=1}^H \pi(\theta_i)$, where $H$ is the number of parameters in $\bm{\theta}$. 

\subsubsection{SPDE formulation}

In the SPDE formulation of the problem, the latent field $\bm{\omega}_t$ is discretized on a mesh with $D$ vertices as described in equation \eqref{basisrep}. Denoting by $\bm{\omega}_t^D$ the vector of values at the nodes of the mesh at time $t$, then $\bm{\omega}_t^D \sim  N(\bm{0}, \bm{Q}_s^{-1})$ where $\bm{Q}_s$ is the precision matrix whose values are computed using \eqref{computeQ}. Details of the computation can be found in \cite{spde}. $\bm{Q}_s$ is a sparse precision matrix whose matrix inverse provides a good approximation of the variance-covariance matrix with elements computed using the Mat\`{e}rn covariance function. This gives 
\begin{align}
	\bm{\xi}_t^D = \varsigma \bm{\xi}_{t-1}^D + \bm{\omega}_t^D, \;\;\; \bm{\omega}_t^D \sim N(\mathbf{0},\textbf{Q}_s^{-1}), \;\;\; t=1,\ldots,T,
\end{align}
where $\bm{\xi}_1^D \sim N\big(\mathbf{0}, \tfrac{1}{1-\varsigma^2} \textbf{Q}_s^{-1}\big)$ and $\bm{\xi}_t^D$ is the vector of values of the spatio-temporal random effect at the nodes of the mesh at time $t$. Since $\textbf{Q}_s$ is a sparse matrix, then the model is a GMRF. The joint distribution of the $TD$-dimensional GMRF $\bm{\xi}^D = \begin{pmatrix}  \bm{\xi}^{D \intercal}_1 & \ldots & \bm{\xi}^{D \intercal}_T  \end{pmatrix}^\intercal$ is  
\begin{align}
	\bm{\xi}^D \sim N(\bm{0}, (\bm{Q}_s \otimes \bm{Q}_T)^{-1}), 
\end{align}
where $\bm{Q}_T$ is the precision matrix for the autoregressive process of order 1, the form of which is given in \cite{rueheld}. 

Since the spatial field is estimated in a mesh whose nodes may be different from the observed locations and from the centroids of the proxy data, then there needs to be a linear mapping from the nodes to the locations of the observed values. This is done by introducing a projection or mapping matrix, say $\bm{B}$, which is a sparse $(M+ G) \times D$ matrix, so that 
\begin{equation}
    \bm{x}_t = \beta_{0}\bm{1}_{G+M} + \beta_{1}\bm{z}_t+ \bm{B} \bm{\xi}_t^D, t=1,\ldots,T
\end{equation}
or
\begin{align}
	x(\textbf{s}_i,t) =  \beta_{0}+ \beta_{1}z(\textbf{s}_i,t)+ \sum_{d=1}^D b_{id} \xi_{td},
\end{align}
where $\xi_{td}$ is the $d$th element of the vector $\bm{\xi}_t^D$ and $b_{id}$ is the $d$th element of the $i$th row of the mapping matrix $\bm{B}$. For the case when all grid centroids and locations of monitors are used as nodes, $\bm{B}$ has only one non-zero and unit element for each row. However, when the resolution of the proxy data is high, the mesh could be too fine and estimation could be more computationally costly.  Hence, the mesh is usually made coarser than the resolution of the proxy data. For this case, many of the elements of $\bm{x}_t$ are different from the nodes, so that their values are now computed via linear interpolation still via the mapping matrix $\bm{B}$. Either way, the matrix $\bm{B}$ remains sparse. 

Hence, \eqref{pseudozeroes} can be written as
\begin{align}
	\bm{0}_t &= -\bm{x}_t+ \beta_{0}\bm{1}_{G+M} + \beta_{1}\bm{z}_t+ \bm{B} \bm{\xi}_t^D 
\end{align}
The first-stage hierarchical SPDE model is then given by
\begin{align}
	\bm{w}_t &= \bm{x}^*_{t,M} + \bm{e}_t, \;\;\; \bm{e}_t\sim N(\textbf{0},\sigma^2_{e}\mathbb{I}_M), \;\; t=1,\ldots, T \\\
	\tilde{\bm{x}}_t &= \alpha_0\bm{1}_G + \bm{x}^{*}_{t,P} + \bm{\delta}_t,\;\;\;\bm{\delta}_t\sim N(\mathbf{0},\sigma^2_{\delta} \mathbb{I}_G), \;\; t=1,\ldots, T \\
\bm{0}_t &= -\begin{pmatrix} \bm{x}_{t,M} \\ \bm{x}_{t,P} \end{pmatrix} + \beta_{0}\mathbf{1}_G + \beta_{1}\bm{z}_t+ \bm{B} \bm{\xi}_t^D \\				
\bm{\xi}_t^D &= \varsigma \bm{\xi}_{t-1}^D + \bm{\omega}_t^D, \;\;\; \bm{\omega}_t^D \sim N(\bm{0},\bm{Q}_s^{-1}), \;\;\; t=1,\ldots,T
\end{align}

The latent Gaussian vector is given by $\begin{pmatrix} \bm{x}^{\intercal} &\bm{x}^{*\intercal} & \bm{\xi}^{\intercal} & \alpha_0 & \beta_0 & \beta_1 \end{pmatrix} ^\intercal$. The vector of hyperparameters is $\bm{\theta} = \begin{pmatrix}  \alpha_1 & \sigma^2_e & \sigma^2_{\delta} & \varsigma & \sigma^2_{\omega} & \kappa \end{pmatrix}^\intercal.$ The model is a latent Gaussian model and hence can be fitted using the INLA method. The posterior marginals of the latent Gaussian vector are computed using \eqref{latentmarg}, while the posterior marginals of the hyperparameters are computed using \eqref{thetamarg}.

\subsubsection{Computing block level exposures}

It is straightforward to predict the values of $x(\bm{s},t)$ on an arbitrary point $\bm{s}$ and at time $t$ using the first-stage hierarchical model. Given the predictions on a grid, two methods to compute the block-level exposures are discussed in \cite{inlaspde}. Suppose we denote by $\hat{x}(B_i,t)$ the predicted value of exposures at block $B_i$ at time $t$. The first method considers all the prediction grids which overlaps with block $B_i$. Suppose we denote by $\hat{x}(G_{\bm{s}_j}, t)$ the predicted value at the grid whose centroid is $\bm{s}_j$ at time $t$ and $h(G_{\bm{s}_j},B_i)$ the proportion of the area of $B_i$ which overlaps with $G_{\bm{s}_j}$. Then method one is computed as follows:
				\begin{align}\label{method1}
					\text{\textbf{Method 1}:} \;\;\;\;\;\;\;\;\;\; \hat{x}(B_i,t) = \sum_{\forall j \ni  G_{\bm{s}_j} \cap B_i \neq \phi} \hat{x}(G_{\bm{s}_j}, t) h(G_{\bm{s}_j},B_i).
				\end{align}
In other words, the value of $\hat{x}(B_i,t)$ is computed as a weighted mean of the predicted values at the grids of the prediction grid which overlaps with block $B_i$, and with weights as the overlap of each grid and the block $B_i$. On the other hand, method 2 uses only the grids whose centroids are located inside block $B_i$. The computed value of $\hat{x}(B_i,t)$ is then computed as a simple mean, i.e., it is computed as 
				\begin{align}\label{method2}
					\text{\textbf{Method 2}:} \;\;\;\;\;\;\;\;\;\; \hat{x}(B_i,t)  = \dfrac{1}{N(\textbf{s}_j \in B_i)} \sum_{\textbf{s}_j \in B_i} \hat{x}(G_{\bm{s}_j}, t). 
				\end{align}

\subsection{Model fitting for second-stage model}
\label{subsec:proposed.stage2.model}

The second stage fits the health model using $\hat{x}(B_i,t)$ as the estimate for $x(B_i,t)$, i.e.,
        \begin{align}
					\hat{x}(B_i,t) \approx \dfrac{1}{|B_i|}\int_{\bm{s} \in B_i} x(\bm{s},t)d\bm{s}.
				\end{align}
The fitting procedure for the second-stage model is more straightforward than the first-stage model. The latent field is given by $\begin{pmatrix} \gamma_0 & \gamma_1 & \bm{\phi}^\intercal & \bm{\nu}^\intercal \end{pmatrix} ^\intercal$ while the hyperparameter vector is $\begin{pmatrix} \sigma^2_{\phi} & \sigma^2_{\nu} \end{pmatrix} ^\intercal$. The posterior distribution of interest is
        \begin{align}
        \pi(\gamma_0,\gamma_1,\bm{\phi},\bm{\nu},\sigma^2_{\phi},\sigma^2_{\nu}|\bm{y}) \propto \pi(\bm{y}|\gamma_0,&\gamma_1,\bm{\phi},\bm{\nu}) \pi(\bm{\phi}|\sigma^2_{\phi}) \pi(\bm{\nu}|\sigma^2_{\nu}) \pi(\gamma_0,\gamma_1,\sigma^2_{\phi},\sigma^2_{\nu}).
	\end{align}
$\pi(\bm{y}|\gamma_0,\gamma_1,\bm{\phi},\bm{\nu})$ has a simple form since conditional on the latent field and hyperparameters, the elements of $\bm{y}$ are independent, i.e., 
\begin{align}
	&Y(B_i,t)|\gamma_0,\gamma_1,\gamma_2,\bm{\phi},\bm{\nu} \stackrel{ind}{\sim} \text{ Poisson}\bigg(\mu(B_i,t)=P(B_i,t) \lambda(B_i,t) \bigg)\\
	&\log(\lambda(B_i,t)) = \gamma_0 + \gamma_1 x(B_i,t) + \phi_i + \nu_t.
\end{align}

Moreover, the posterior distribution of $\bm{\phi}$ is trivial since $\bm{\phi} \sim N(\bm{0},\sigma^2_{\phi} \mathbb{I})$. The joint distribution of $\bm{\nu} = \begin{pmatrix} \nu_1 & \ldots &  \nu_T \end{pmatrix}$ uses the fact that the increments are independent, i.e., $\nu_{t+1}-\nu_t \sim N(0,\sigma^2_{\nu}).$ Since the second-stage model is a latent Gaussian model, the posterior marginals of the latent field $\begin{pmatrix} \gamma_0 & \gamma_1 & \gamma_2 & \bm{\phi}^\intercal & \bm{\nu}^\intercal \end{pmatrix} ^\intercal$ and the hyperparameter space $\begin{pmatrix} \sigma^2_{\phi} & \sigma^2_{\nu} \end{pmatrix} ^\intercal$ can be approximated with INLA using equations \eqref{latentmarg} and \eqref{thetamarg}.

\subsubsection{Propagating uncertainty from stage 1 to stage 2} \label{subsec:propagate.uncertainty}

One approach to propagate uncertainty from stage 1 to stage 2 is to sample several times from the posterior predictive distribution of $\bm{x}_t$ from the first-stage model, and then to compute the exposures at the block level from each sample using $\eqref{method1}$ and $\eqref{method2}$ which are then fed into the stage 2 model \cite{inlaspde, BLANGIARDO20161, LeeMukhopadhyayetal, liu2017incorporating}. The final parameter estimates of the second-stage model are computed using the combined results from all samples. The steps are as follows:
 
Do $i=1,\ldots,J$:
		\begin{enumerate}
			\item Simulate from the marginal posterior distribution of the latent field $\hat{\pi}(x(\bm{s},t)|\cdot)$ on the prediction grid. 
			\item Compute block-level exposures, $\hat{x}(B_i,t)$, using the two methods in equations $\eqref{method1}$ and $\eqref{method2}$.
			\item Fit the stage 2 GLMM using INLA as described in section \ref{subsec:proposed.stage2.model}.
			\item For each parameter in the stage 2 model, generate samples from its posterior marginal distribution. 
		\end{enumerate}
After completing all $J$ cycles from above, all samples from step (4) are combined and used to approximate the posterior distribution of the stage 2 parameters and also to compute point estimates and interval estimates.

\section{Simulation Study} \label{sec:simulation.study}

The performance of the proposed method is investigated using a simulation study. The study region used is the Belo Horizonte region in Brazil which is available in the \textit{spdep} package in R  \cite{bivand2015comparing} and is the same study region used in \cite{inlaspde}. Figure \ref{fig:sparseVSnonsparse} shows the study region map. The shapefile has a total of 98 blocks or areas.

\subsection{Simulation of the exposure field}

For the Mat\`{e}rn covariance function parameters, the spatial variance $\sigma^2_{\omega}$ is 1.5, while the range parameter $\rho$ is 1.89 which corresponds to around 46\% of the maximum distance in the $100\times100$ simulation grid. The autoregressive parameter $\varsigma$ is set to 0.7. The single covariate $z(\bm{s},t)$ was generated from $N(\mu=0,\sigma=1)$. The fixed effects are $\beta_0=0$ and $\beta_1=2$. The simulated values of $x(B_i,t)$ is then a spatial average of all the points inside block $B_i$, i.e.,
        \begin{align}
		x(B_i,t) = \dfrac{1}{|B_i|}\int_{B_i} x(\bm{s},t)d\bm{s} \approx \sum_{\forall \bm{s} \in B_i} \dfrac{1}{\#(\bm{s} \in B_i)}x(\bm{s},t).
	\end{align}

\subsection{Simulation of the health data}

Using the simulated values of $x(B_i,t)$, it will be straightforward to simulate the health data. The fixed effects are $\gamma_0=-3$ and $\log(\gamma_1)=1.2.$ The assumed value of $\gamma_1$ implies an expected increase of 20\% in the relative risk for a one unit increase in $x(B_i,t)$. The assumed values of the variance parameters of the block-specific effect and the time effect are $\sigma^2_{\phi} = \sigma^2_{\nu} = 0.02$. The values of the expected values for each block are generated from a uniform distribution and are designed to be proportional to the size of the block so that blocks with bigger surface areas have higher expected values. 

\subsection{Simulation of the monitors and proxy data}

\begin{figure}[h]
        \centering
     	\includegraphics[trim={0 0 0 3cm},clip,width=.7\textwidth]{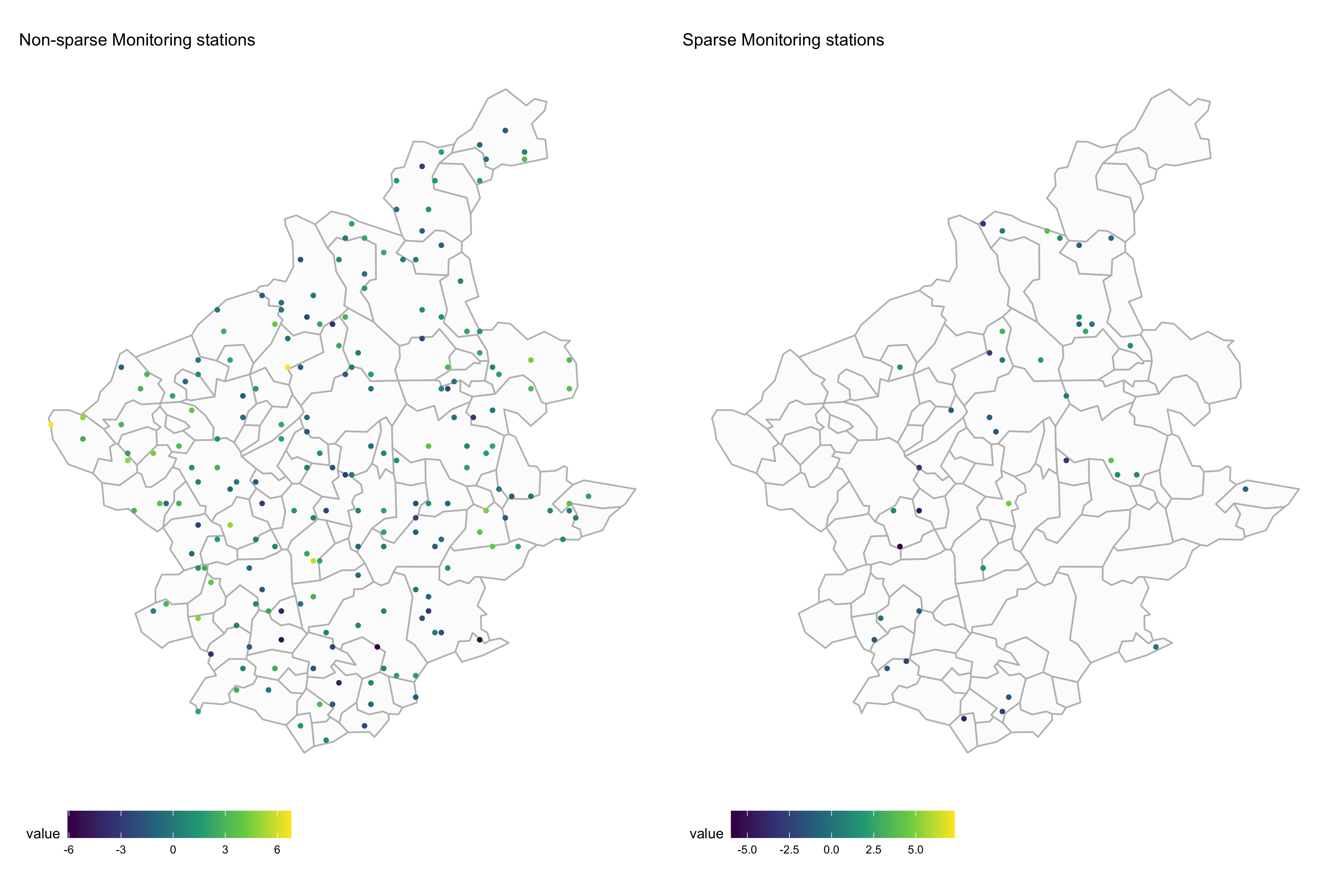}
	\caption{Non-sparse network of monitors (left) and a sparse network of monitors (right)}
	\label{fig:sparseVSnonsparse}
\end{figure}

\begin{figure}[!ht]
        \centering
     	\includegraphics[trim={0 2cm 0 2cm},clip,width=.7\textwidth]{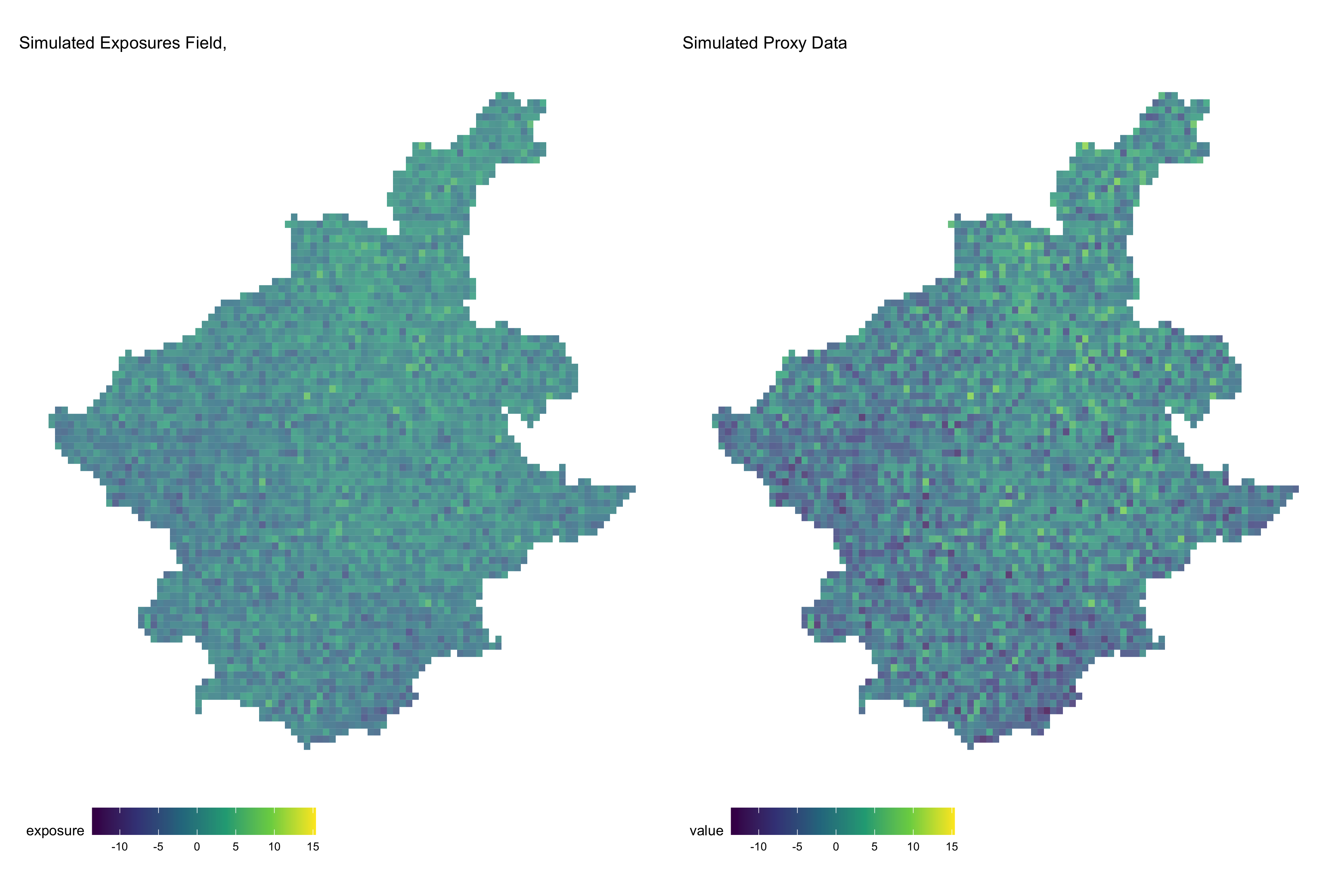}
	\caption{Simulated true exposures field (left) and corresponding proxy data after adding biases (right)}
	\label{fig:exposuresVSproxydata}
\end{figure}

The monitors are simulated by getting a random sample of points from the simulation grid. A non-sparse network was considered and investigated in \cite{inlaspde} - either getting 2\%, 10\%, or 30\% of the points from the simulation grid inside each block. Their simulation results showed that area predictions are better in terms of the RMSE and the correlation with the true block-level exposure values when there are more monitoring stations in the data. A contribution of this paper is to look at the case of having a sparse network of monitoring stations, i.e., there are few monitors and several areas or blocks don't have monitors inside. This study considers two scenarios for the monitoring stations data. The first scenario is a non-sparse network similar to that considered in \cite{inlaspde}. The second scenario is a sparse network and is carefully chosen in such a way that it resembles how actual data of sparse monitoring stations look like. Figure \ref{fig:sparseVSnonsparse} shows an example of a non-sparse network (left) and a sparse network (right) of monitors. The known values $w(\bm{s}_i,t), i=1,\ldots, M$ follow the classical error model, with the error term assumed as $e(\bm{s}_i,t) \sim N(0,\sigma^2_e=0.1).$
The simulation of the proxy data incorporates additive and multiplicative biases given by $\alpha_0= -1, \alpha_2=1.5$, and an additional Gaussian noise term $\bm{\delta_t} \sim N(\bm{0},\sigma^2_{\delta}\mathbb{I}))$ where $\sigma^2_{\delta} = 1.$ Figure \ref{fig:exposuresVSproxydata} shows a sample exposures field (left) and the known proxy data (right) after adding the biases at a fixed time. The noise is very apparent in the proxy data. 

\subsection{Prediction grid}

The effect of the resolution of the prediction grid on the block-level predictions using equations \eqref{method1} and \eqref{method2}  has been investigated in \cite{inlaspde}. Their simulation results have shown that with a finer prediction grid, the block-level exposure predictions are also more accurate. Hence, in the simulation study in this paper, a $100 \times 100$ prediction is used for all the scenarios, which is the same grid resolution used to simulate the true exposures field and the resolution of the proxy data. 

\subsection{Simulation scenarios}

\begin{figure}[!ht]
        \centering
     	\includegraphics[trim={0 0cm 0 0cm},clip,width=.7\textwidth]{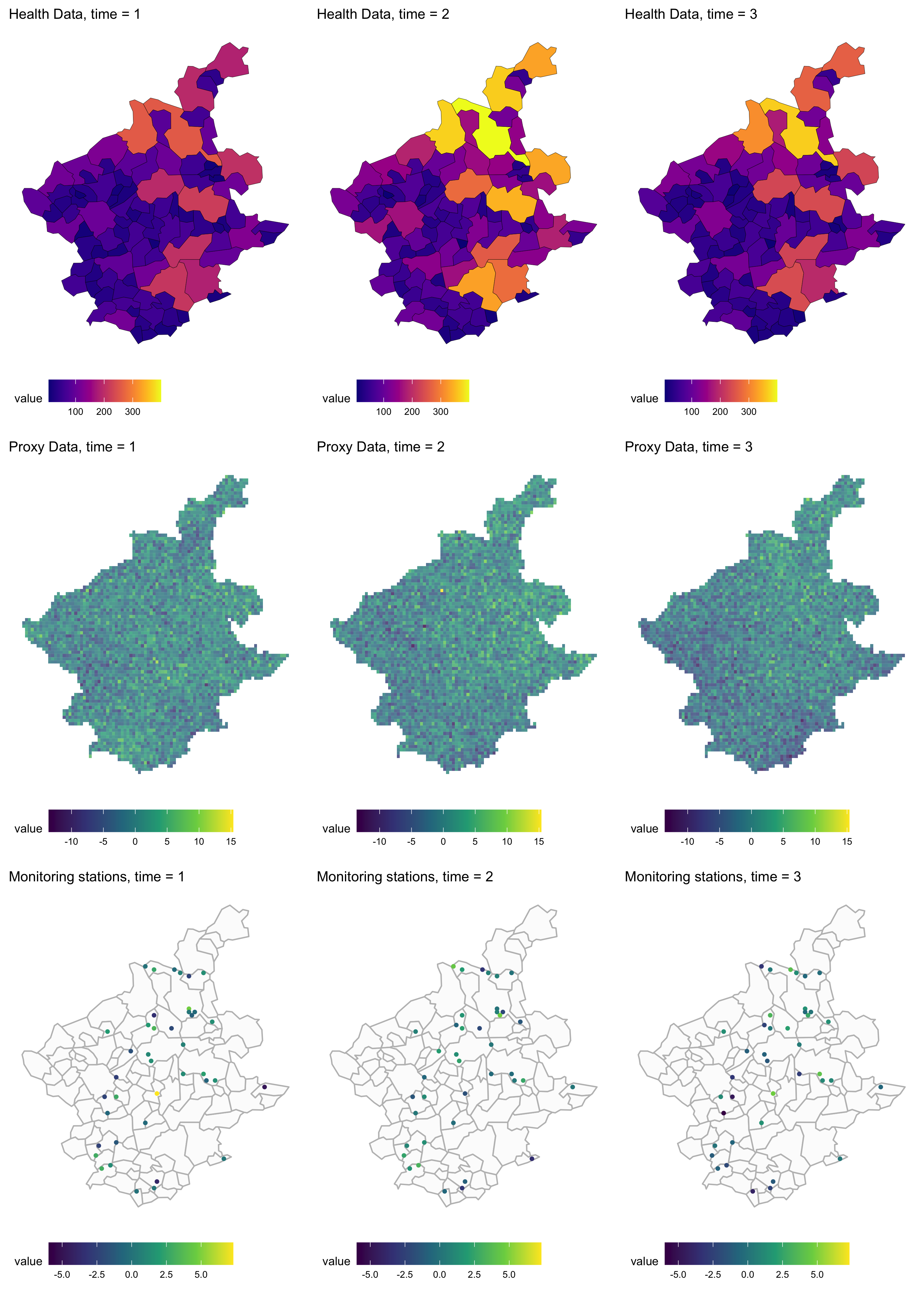}
	\caption{Sample data for a spatio-temporal analysis: count data (top row), proxy data (middle row), monitoring stations data (bottom)}. 
	\label{fig:DataExample}
\end{figure}

Figure \ref{fig:DataExample} shows a sample data with $T=3$ for a spatio-temporal analysis: the counts at the blocks which is assumed to be Poisson (top), the proxy data which is a high resolution data on a regular grid (middle), and data from a network of monitoring stations (bottom). 
In addition to this, there are also additional data on a covariate for the stage-two model and another covariate for the first-stage model at the level of the monitors and the proxy data. There are three simulation settings considered in the study:
\begin{enumerate}
	\item The sparsity of the monitoring stations: sparse or non-sparse. This is illustrated in Figure \ref{fig:sparseVSnonsparse}. 
	\item Length of time: $T=3$, $T=6$, or $T=12$. Since this simulation study is a spatio-temporal extension of that in \cite{inlaspde}, it is important to investigate the effect of the length of time in terms of the model estimates and predictions.
	\item Prior specification: use of non-informative priors or informative priors.  In Bayesian analysis, sensitivity of the posterior estimates is an important part of the analysis. The effect of prior specification is therefore investigated in this simulation study. Not all parameters are given informative priors. The parameters provided with informative priors are those which are usually difficult to estimate which include the parameters of the latent Gaussian field $\sigma^2_{\xi}, \rho$, and $\varsigma$; the variance parameters $\sigma^2_{\nu}, \sigma^2_{\phi}, \sigma^2_{e},$ and $\sigma^2_{\delta}$; and $\alpha_1$ which is the scaling parameter of $\bm{x}_{t,P}$.
\end{enumerate}
	
There are a total of 12 simulation scenarios. Table \ref{tab:sim.scenarios} shows all the simulation scenarios and how they are labelled in the figures in the results section and in the appendix. For each scenario, 500 independent replications are done in order to evaluate the performance of the proposed method. The metrics used to evaluate the performance of the proposed method are discussed in section \ref{subsub:model.eval}. Also, in fitting the second-stage model, the number of simulations from the posterior predictive distribution of the latent field $x(\bm{s},t)$ is set at $J=100$. For each posterior marginal, 200 random values were generated to compute posterior quantities of interest which includes the posterior mean, posterior median, and 95\% credible intervals.

\begin{table}[ht]
\centering
\resizebox{0.55\linewidth}{!}{
\begin{tabular}{|c|c|lll|}
\hline
\multirow{2}{*}{\textbf{Sparsity}} & \multirow{2}{*}{\textbf{Prior Specification}} & \multicolumn{3}{c|}{\textbf{Time}}                      \\ \cline{3-5} 
                                   &                                                & \multicolumn{1}{l|}{3} & \multicolumn{1}{l|}{6} & 12 \\ \hline
\multirow{2}{*}{No}                & Informative                                    & \multicolumn{1}{l|}{A} & \multicolumn{1}{l|}{E} & I  \\ \cline{2-5} 
                                   & Non-informative                                & \multicolumn{1}{l|}{B} & \multicolumn{1}{l|}{F} & J  \\ \hline
\multirow{2}{*}{Yes}               & Informative                                    & \multicolumn{1}{l|}{C} & \multicolumn{1}{l|}{G} & K  \\ \cline{2-5} 
                                   & Non-informative                                & \multicolumn{1}{l|}{D} & \multicolumn{1}{l|}{H} & L  \\ \hline
\end{tabular}}
\caption{\label{tab:sim.scenarios}Simulation scenarios in the simulation study}
\end{table}

\subsubsection{Specification of Non-informative Priors}

For the case when non-informative priors are used, the following priors are used for the first-stage parameters: $\alpha_0 \sim N(0,\infty)$, $\beta_0 \sim N(0,\infty)$, $\alpha_1 \sim N(0,1000)$, $\beta_1 \sim N(0,1000)$, $\tfrac{1}{\sigma^2_e} \sim \text{Gamma}(1,0.00005)$, $\tfrac{1}{\sigma^2_{\delta}} \sim \text{Gamma}(1,0.00005)$. The priors of the Mat\`{e}rn covariance function are defined in terms of $\log(\tau)$ and $\log(\kappa)$, both of which are functions of the SPDE parameters $\sigma^2_{\omega}, \nu$, and $\rho$. It is assumed that both $\log(\tau)$ and $\log(\kappa)$ are jointly Gaussian. A detailed discussion of prior specification for Mat\`{e}rn models using the SPDE approach is discussed in \cite{lindgren2015bayesian}. The non-informative prior for the AR(1) parameter is defined in terms of $\log\bigg( \dfrac{1+\varsigma}{1-\varsigma} \bigg)$, specified as $\log\bigg( \dfrac{1+\varsigma}{1-\varsigma} \bigg) \sim N(0, 0.15^2)$. For the second-stage model parameters, the non-informative priors for the fixed effects are $\gamma_0 \sim N(0,\infty)$ and $\gamma_1 \sim N(0,1000)$. The variance parameters have the following non-informative priors: $\tfrac{1}{\sigma^2_{\phi}} \sim \text{Gamma}(1,0.00005)$ and $\tfrac{1}{\sigma^2_{\nu}} \sim \text{Gamma}(1,0.00005)$.

\subsubsection{Specification of Informative Priors}

The specification of priors for the parameters of the Gaussian random field, $\sigma^2_{\omega}$ and $\rho$, is more carefully considered because estimating models which include a GRF may lead to overfitting \cite{fuglstad2019constructing}. For the case of having informative priors, the so-called \textit{penalized complexity priors} (PC priors) proposed in \cite{fuglstad2019constructing} are used. The PC prior is a weakly informative prior which penalizes complexity or additional flexibility in the model; hence, the prior tends to prefer the simpler \textit{base} model. It works on the principle that a model further away from the base model should be more strongly penalized, and that the penalty is of constant rate with respect to the distance between the two models as measured by the \textit{Kullback-Leibler divergence}. In the context of a GRF with Mat\`{e}rn covariance function, the PC prior shrinks the model to the base model with infinite range and zero marginal variance. The implementation of the PC priors for $\sigma^2_{\omega}$ and $\rho$ works on a reparametrized version of the Mat\`{e}rn covariance function, which holds in dimension not more than three and under the following conditions:
\begin{align}
	\mathbb{P}(\sigma_{\omega} > \sigma_{\omega,0} | \kappa) &= \alpha \\
	\mathbb{P}(\rho < \rho_0 ) & = \alpha
\end{align}
The conditions above make it possible to set the priors in interpretable terms, i.e.\ in terms of the marginal standard deviation and the range parameter. The conditions above are a joint specification of the priors, where $\sigma_{\omega,0}$ and $\rho_0$ are the upper and lower limit for $\sigma_{\omega}$ and $\rho$, respectively, and $\alpha$ is the tail probability. In fitting the models, $\sigma_{\omega,0}$ and $\rho_0$ are set as equal to the true value, and $\alpha = 0.05$. For the measurement error variance parameters, $\sigma^2_{\delta}$ and $\sigma^2_{e}$, the prior is inverse-Gamma with mean close to the true value and a small variance. For the scaling parameter $\alpha_1$, the prior is Gaussian with mean equal to the true value and a small variance. For the AR(1) parameter, the prior is also defined on the same transform and is Gaussian with mean equal to the true value and with a small variance. The other parameters in the first-stage model are given the same non-informative priors.  

For the second-stage model parameters, the precision parameters $\sigma^2_{\phi}$ and $\sigma^2_{\nu}$ are given the PC priors, which are specified as $\mathbb{P}(\sigma_{\phi}>\sigma_{\phi,0}) = \alpha$ and $\mathbb{P}(\sigma_{\nu}>\sigma_{\nu,0}) = \alpha$, where $\sigma_{\phi,0}$ and $\sigma_{\nu,0}$ are set as equal to its true value and $\alpha = 0.05.$ Finally, the fixed effects $\gamma_0$ and $\gamma_1$ are given the same non-informative priors. 

There are three precision parameters in the first-stage model that are fixed at an appropriate level. The first one is the precision parameter, $\tau_0$, for the pseudo-zeroes which is fixed at a large value. The second one is the precision parameter, $\tau_x$, for the prior of the latent field $\bm{x}_t$ which is fixed at a very small value. Finally, for the conditional distribution of the copies $\bm{x}_{t,M}^*$ and $\bm{x}_{t,P}^*$, the precision parameter $\tau_x*$ is also fixed at a large value so that both mimic $\bm{x}_{t,M}$ and $\bm{x}_{t,P}$, respectively.

\subsubsection{Model evaluation} \label{subsub:model.eval} 
 
The performance of the estimation procedure are evaluated in terms of the bias, RMSE, and coverage probability for each parameter estimate and block-level prediction $\hat{x}(B_i,t)$. There is a slight difference in the notations for the formulas to compute the model performance metrics between the first-stage and second-stage model parameters. For a first-stage parameter $\theta$, $\hat{\theta}_{ik}$ denotes the $k$th sampled value from the estimated posterior distribution of $\theta$ in the $i$th replication, $i=1,\ldots, n_{sim}$ and $k=1,\ldots,K.$ For a second-stage parameter $\theta$, $\hat{\theta}_{ijk}$ denotes the $k$th sampled value from the estimated posterior distribution of $\theta$ using the block-level estimates $\hat{x}(B,t)$ computed using the $j$th simulated values from the marginal posterior distribution of the latent field, $j=1,\ldots,J$, $k=1,\ldots,K$, $i=1,\ldots,n_{sim}$. In the simulation study, $n_{sim}=500, J=50$ and $K=200$. 
\begin{enumerate}
	\item \textbf{Bias} - The error in the estimation for the $i$th replicate is simply the mean of the deviation from the true value $\theta$ for each sampled value $\hat{\theta}_{ik}, k=1,\ldots,K$ from its posterior distribution. Denoting by $\text{error}_i$ the error in the $i$th replicate, we have
	\begin{align}
	\text{error}_i = \dfrac{1}{K}\sum_{k=1}^K (\hat{\theta}_{ik}-\theta)=\dfrac{1}{K}\sum_{k=1}^K \hat{\theta}_{ik} -\theta = \hat{\theta}_i-\theta,
	\end{align}
where $\hat{\theta}_i$ is the estimated value of $\theta$ at the $i$th replication. The bias is then computed as the average of $\text{error}_i$ for all $i$, i.e.,
	\begin{align}
		\text{bias} = \dfrac{1}{n_{sim}} \sum_{i=1}^{n_{sim}}\text{error}_i =  \dfrac{1}{n_{sim}} \sum_{i=1}^{n_{sim}} \hat{\theta} - \theta
	\end{align}
A slight modification in the formula for $\text{error}_i$ for a second-stage parameter is as follows:
	\begin{align}
	\text{error}_i = \dfrac{1}{JK}\sum_{j=1}^J\sum_{k=1}^K (\hat{\theta}_{ijk}-\theta) = \hat{\theta}_i-\theta.
	\end{align}
The bias is then computed similarly as above, which is simply the average of $\text{error}_i$ for all $i=1,\ldots,n_{sim}$.

	\item \textbf{Coverage probability} - For a given replicate $i$, suppose $\hat{\theta}_i^{(2.5)}$ is the 2.5th percentile of $\hat{\theta}_{ik}$ for all $k$ for a first-stage parameter or of $\hat{\theta}_{ijk}$ for all $j, k$ for a second-stage parameter. Similarly, suppose $\hat{\theta}_i^{(97.5)}$ is the 97.5th percentile. Defining the indicator function
	\begin{align}
		\mathbb{I}_i(\theta) = \begin{cases} 
      1 & \hat{\theta}_i^{(2.5)} < \theta < \hat{\theta}_i^{(97.5)} \\
      0 & \text{otherwise}
   \end{cases},
	\end{align}
the 95\% coverage probability for $\theta$ is then the average of $\mathbb{I}_i(\theta)$ for all replicates, i.e.,
	\begin{align}
	\text{coverage probability} =\dfrac{1}{n_{sim}} \sum_{i=1}^{n_{sim}}\mathbb{I}_i(\theta). 
	\end{align}

	\item \textbf{Root mean square error (RMSE)} - The RMSE is computed as 
	\begin{align}
		\text{RMSE} =  \dfrac{1}{n_{sim}} \sum_{i=1}^{n_{sim}} \sqrt{\dfrac{1}{K} \sum_{k=1}^{K} (\hat{\theta}_{ik} - \theta)^2 }
	\end{align}
for a first-stage parameter. A slight modification for a second-stage parameter is as follows:
	\begin{align}
		\text{RMSE} =  \dfrac{1}{n_{sim}} \sum_{i=1}^{n_{sim}} \sqrt{\dfrac{1}{JK} \sum_{j=1}^{J}\sum_{k=1}^{K} (\hat{\theta}_{ijk} - \theta)^2 }
	\end{align}
\end{enumerate}

The same model evaluation metrics are used for the block-level exposure estimates. A slight modification in the formula is to account for the fact that the true values of the block-level exposures vary for the different replicates. 

\section{Results and Discussion} \label{sec:results.discuss}

The section presents the results of the simulation study. Section 5.1 presents the results for the exposure model parameters, section 5.2 for the health model parameters, and section 5.3 for the block-level exposure estimates. 

\subsection{First-stage model parameters}

\begin{table}[ht]
\resizebox{\linewidth}{!}{%
\begin{tabular}{|c|c|c|c|c|c|c|c|c|c|c|c|}
\hline
\textbf{T}          & \textbf{Sparse}      & \textbf{Priors}  & $\bm{\beta_0}$ & $\bm{\beta_1}$ & $\bm{\alpha_0}$ & $\bm{\alpha_1}$ & $\bm{\sigma_{\delta}^2}$ & $\bm{\sigma_{e}^2}$ & $\bm{\sigma_{\omega}^2}$ & $\bm{\rho}$ & $\bm{\varsigma}$ \\ \hline
                     &                       & informative     &  -0.0066 & -0.0030 & 0.0005 & 0.0025 & 0.0060 & -0.0003 & 0.1217 & 0.1545 & -0.0158  \\ \cline{3-12} 
                     & \multirow{-2}{*}{No}  & non-informative & -0.0081                                     & -0.0022                                     & 0.0004                                      & 0.0017                                      & 0.0052                                      & -0.0024                                     & -0.164                                      & -0.0813                                     & -0.0254                                     \\ \cline{2-12} 
                     &                       & informative     & 0.018                                       & -0.0024                                     & -0.0073                                     & 0.0029                                      & 0.0064                                      & -0.0002                                     & 0.0919                                      & 0.1026                                      & -0.0184                                     \\ \cline{3-12} 
\multirow{-4}{*}{3}  & \multirow{-2}{*}{Yes} & non-informative & 0.0171                                      & 0.0016                                      & -0.0074                                     & -0.0001                                     & 0.0062                                      & -0.0069                                     & 0.7188                                      & 0.3839                                      & -0.0246                                     \\ \hline
                     &                       & informative     & 0.004                                       & -0.0027                                     & 0.0002                                      & 0.0022                                      & 0.0069                                      & 0.0014                                      & 0.056                                       & 0.1346                                      & -0.02                                       \\ \cline{3-12} 
                     & \multirow{-2}{*}{No}  & non-informative & 0.0036                                      & -0.002                                      & -0.0001                                     & 0.0016                                      & 0.0063                                      & -0.0016                                     & -0.1876                                     & -0.0869                                     & -0.0219                                     \\ \cline{2-12} 
                     &                       & informative     & -0.0073                                     & -0.0023                                     & 0.0025                                      & 0.0023                                      & 0.0055                                      & 0.0024                                      & 0.037                                       & 0.0726                                      & -0.0178                                     \\ \cline{3-12} 
\multirow{-4}{*}{6}  & \multirow{-2}{*}{Yes} & non-informative & -0.0091                                     & -0.001                                      & 0.0023                                      & 0.0012                                      & 0.0057                                      & -0.0056                                     & 0.3567                                      & 0.2411                                      & -0.012                                      \\ \hline
                     &                       & informative     & 0.0044                                      & -0.0027                                     & -0.0011                                     & 0.0019                                      & 0.0062                                      & 0.0025                                      & 0.0037                                      & 0.1127                                      & -0.023                                      \\ \cline{3-12} 
                     & \multirow{-2}{*}{No}  & non-informative & 0.0074                                      & -0.002                                      & -0.0011                                     & 0.0013                                      & 0.0055                                      & -0.0023                                     & -0.2501                                     & -0.1247                                     & -0.0212                                     \\ \cline{2-12} 
                     &                       & informative     & 0.0022                                      & -0.0011                                     & 0.0002                                      & 0.0014                                      & 0.0058                                      & 0.0051                                      & 0.019                                       & 0.0662                                      & -0.0217                                     \\ \cline{3-12} 
\multirow{-4}{*}{12} & \multirow{-2}{*}{Yes} & non-informative & 0.0061                                      & -0.0023                                     & 0.0001                                      & 0.0022                                      & 0.006                                       & -0.0032                                     & 0.3978                                      & 0.2814                                      & -0.0073                                     \\ \hline
\end{tabular}}
\caption{\label{tab:bias.exposures}Biases of exposure model parameters for all scenarios}
\end{table}

Table \ref{tab:bias.exposures} shows the biases of the exposure model parameters while Table \ref{tab:rmse.exposures} shows the RMSEs computed from the 500 replicates. For the fixed effects $\beta_0$ and $\beta_1$, the biases are close to zero for all the scenarios. However, when the data on the monitors is sparse, the RMSE of $\beta_1$ is high. The parameter $\beta_1$ is the coefficient of the covariate of the latent exposures field. Although this information is also available for the high-resolution proxy data, its values are noisy and are less correlated with the true values of the latent field. Hence, this could be the reason for the difficulty in estimating correctly $\beta_1$ when the monitors data is sparse, even with a covariate information for the proxy data. But for more time points, the RMSEs are generally smaller. The same pattern can be seen for the intercept $\beta_0$.

For the Mat\`{e}rn field parameters $\sigma^2_{\xi}$ and $\rho$, the sparsity of the monitors data and the prior specification has a potential effect on the quality of the estimates. The biases are closer to zero and the RMSEs are generally lower when penalized complexity priors are used and when the data from monitors is not sparse. Even if the monitors data is sparse, the biases and RMSEs are still small as long as penalized complexity priors are used. This is expected since if there are few observed values from monitors, we would rely on informative priors to correctly estimate the Mat\`{e}rn parameters. When non-informative priors are used and data on the monitors is sparse, the bias and RMSEs are expected to be high.

For the AR parameter $\varsigma$, it is mainly the prior specification which determines the quality of the obtained estimates. Although the biases are generally close to zero for all scenarios, the RMSE is generally higher when the priors are non-informative, regardless of the sparsity of the monitors data. Nonetheless, the RMSE of $\varsigma$ is generally smaller when there are more time points. This is expected since $\varsigma$ parametrizes the temporal evolution of the spatial field, so that more time points means more information available to estimate $\varsigma$. 

For the measurement error variance $\sigma^2_e$ in the monitors, the biases are close to zero for all scenarios but the RMSEs are generally higher when the data on the monitors is sparse. This is expected since if there are few data points from monitors, there is also little information at hand to do the estimation. But even if the data on the monitors is sparse, for as long as informative priors are used or there are several time points, the RMSEs are generally lower. For the additive and multiplicative bias in the proxy data, $\alpha_0$ and $\alpha_1$, the biases are generally close to zero for all scenarios. However, the RMSEs are generally higher when the data on the monitors is sparse. Regardless of the sparsity in the monitors data, the RMSEs are generally smaller if there are more time points in the data. Finally, for the error variance of the proxy data, $\sigma^2_{\delta}$, the biases are also close to zero for all scenarios. The RMSEs are also generally small, which decreases for scenarios with more time points.

\begin{table}[ht]
\resizebox{\linewidth}{!}{%
\begin{tabular}{|c|c|c|c|c|c|c|c|c|c|c|c|}
\hline
\textbf{T}          & \textbf{Sparse}      & \textbf{Priors}  & $\bm{\beta_0}$ & $\bm{\beta_1}$ & $\bm{\alpha_0}$ & $\bm{\alpha_1}$ & $\bm{\sigma_{\delta}^2}$ & $\bm{\sigma_{e}^2}$ & $\bm{\sigma_{\omega}^2}$ & $\bm{\rho}$ & $\bm{\varsigma}$ \\ \hline
\multirow{4}{*}{3}  & \multirow{2}{*}{No}  & informative     & 1.0256      & 0.0124      & 0.0325      & 0.0138      & 0.0184              & 0.0078          & 0.5327           & 0.3748         & 0.0541      \\ \cline{3-12} 
                    &                      & non-informative & 0.9053      & 0.0125      & 0.0323      & 0.0137      & 0.0182              & 0.0097          & 0.576            & 0.423          & 0.0617      \\ \cline{2-12} 
                    & \multirow{2}{*}{Yes} & informative     & 1.0107      & 0.0252      & 0.0713      & 0.0301      & 0.0186              & 0.011           & 0.5421           & 0.3684         & 0.0606      \\ \cline{3-12} 
                    &                      & non-informative & 1.2309      & 0.0247      & 0.0697      & 0.0287      & 0.0188              & 0.0218          & 1.2865           & 0.7018         & 0.0699      \\ \hline
\multirow{4}{*}{6}  & \multirow{2}{*}{No}  & informative     & 0.8616      & 0.0091      & 0.0219      & 0.0098      & 0.0145              & 0.0075          & 0.3729           & 0.2825         & 0.0394      \\ \cline{3-12} 
                    &                      & non-informative & 0.7709      & 0.0091      & 0.0218      & 0.0096      & 0.0142              & 0.0073          & 0.4694           & 0.3403         & 0.0434      \\ \cline{2-12} 
                    & \multirow{2}{*}{Yes} & informative     & 0.8639      & 0.0168      & 0.0509      & 0.0205      & 0.0137              & 0.0106          & 0.372            & 0.2595         & 0.0396      \\ \cline{3-12} 
                    &                      & non-informative & 1.0298      & 0.0167      & 0.0496      & 0.0198      & 0.0139              & 0.0158          & 0.7297           & 0.4587         & 0.0451      \\ \hline
\multirow{4}{*}{12} & \multirow{2}{*}{No}  & informative     & 0.6787      & 0.0065      & 0.0165      & 0.0071      & 0.0106              & 0.0068          & 0.2685           & 0.215          & 0.0325      \\ \cline{3-12} 
                    &                      & non-informative & 0.5975      & 0.0065      & 0.0164      & 0.0071      & 0.0103              & 0.0058          & 0.3869           & 0.2703         & 0.0323      \\ \cline{2-12} 
                    & \multirow{2}{*}{Yes} & informative     & 0.6794      & 0.0121      & 0.037       & 0.0145      & 0.0104              & 0.0093          & 0.2616           & 0.1851         & 0.0326      \\ \cline{3-12} 
                    &                      & non-informative & 0.8194      & 0.0127      & 0.0366      & 0.0144      & 0.0106              & 0.0111          & 0.6138           & 0.3951         & 0.0346      \\ \hline

\end{tabular}}
\caption{\label{tab:rmse.exposures}RMSEs of exposure model parameters for all scenarios}
\end{table}

\begin{table}[ht]
\resizebox{\linewidth}{!}{%
\begin{tabular}{|c|c|c|c|c|c|c|c|c|c|c|c|}
\hline
\textbf{T}          & \textbf{Sparse}      & \textbf{Priors}  & $\bm{\beta_0}$ & $\bm{\beta_1}$ & $\bm{\alpha_0}$ & $\bm{\alpha_1}$ & $\bm{\sigma_{\delta}^2}$ & $\bm{\sigma_{e}^2}$ & $\bm{\sigma_{\omega}^2}$ & $\bm{\rho}$ & $\bm{\varsigma}$ \\ \hline
\multirow{4}{*}{3}  & \multirow{2}{*}{No}  & informative     & 95.8                             & 58.6                             & 92.6                             & 95.2                             & 92                                       & 89                                   & 99.8                                  & 98.8                                & 97                               \\ \cline{3-12} 
                    &                      & non-informative & 83.2                             & 56.8                             & 91.6                             & 94                               & 92.2                                     & 90.2                                 & 68.4                                  & 70.8                                & 91.6                             \\ \cline{2-12} 
                    & \multirow{2}{*}{Yes} & informative     & 93.6                             & 28.6                             & 93.2                             & 94.2                             & 91.6                                     & 98.8                                 & 97.2                                  & 97.6                                & 97.6                             \\ \cline{3-12} 
                    &                      & non-informative & 94                               & 31.4                             & 91.4                             & 91.6                             & 88.6                                     & 82                                   & 81                                    & 80.2                                & 85.2                             \\ \hline
\multirow{4}{*}{6}  & \multirow{2}{*}{No}  & informative     & 94.2                             & 56.6                             & 93.6                             & 92.8                             & 85.4                                     & 71.8                                 & 98.6                                  & 96.8                                & 91.8                             \\ \cline{3-12} 
                    &                      & non-informative & 82.6                             & 57.8                             & 93.2                             & 92.8                             & 85.4                                     & 86.6                                 & 61.4                                  & 63.2                                & 87                               \\ \cline{2-12} 
                    & \multirow{2}{*}{Yes} & informative     & 94                               & 32                               & 92.6                             & 94                               & 87                                       & 98.2                                 & 99.2                                  & 97.8                                & 92.4                             \\ \cline{3-12} 
                    &                      & non-informative & 93.8                             & 35                               & 92.6                             & 92.6                             & 87.2                                     & 80.8                                 & 84                                    & 82.4                                & 83                               \\ \hline
\multirow{4}{*}{12} & \multirow{2}{*}{No}  & informative     & 94.4                             & 57.4                             & 93                               & 92.6                             & 82.4                                     & 58.6                                 & 99.4                                  & 93.4                                & 81.8                             \\ \cline{3-12} 
                    &                      & non-informative & 85                               & 57                               & 93                               & 90.4                             & 83.4                                     & 74                                   & 48                                    & 54.4                                & 79.8                             \\ \cline{2-12} 
                    & \multirow{2}{*}{Yes} & informative     & 96                               & 35.6                             & 94.2                             & 93.8                             & 84.6                                     & 92.8                                 & 99.4                                  & 95.2                                & 83.4                             \\ \cline{3-12} 
                    &                      & non-informative & 96.8                             & 32.8                             & 91.6                             & 90                               & 80                                       & 83.4                                 & 78.2                                  & 71.4                                & 73                               \\ \hline

\end{tabular}}
\caption{\label{tab:coverage.exposures}Coverage probabilities of exposure model parameters for all scenarios}
\end{table}

Table \ref{tab:coverage.exposures} shows the coverage probabilities of the exposure model parameters. For the fixed effect $\beta_0$, the coverage probabilities are generally high for all the scenarios. However, for $\beta_1$, the coverage probabilities are surprisingly low especially for the case when the data on the monitors is sparse. Note that Tables \ref{tab:bias.exposures} and \ref{tab:rmse.exposures} show that the biases and the RMSEs for $\beta_1$ are close to zero for all scenarios, so the low coverage probabilities mean that the obtained credible intervals are too narrow to contain the true parameter value, but the point estimates are nonetheless close to the true value. 

For the Mat\`{e}rn field parameters, the coverage probabilities are very high when informative priors are used, which is the expected result. When non-informative priors are used, the coverage probabilities for the case of sparse monitors data is higher compared to the case of non-sparse data. This may seem counter-intuitive, but as show in Table \ref{tab:rmse.exposures}, the RMSEs for $\sigma^2_{\xi}$ and $\rho$ are very large when the monitors data is sparse which implies very wide credible intervals, and that the RMSEs are smaller when the monitors data is non-sparse which implies narrower credible intervals. This could be the explanation for larger coverage probabilities when monitors data is sparse; nonetheless, in terms of the bias and RMSE, the obtained estimates of $\sigma^2_{\xi}$ and $\rho$ are better when monitors data is not sparse. 

For the AR parameter $\varsigma$, the coverage probabilities are generally lower when the priors used are non-informative, which is also the expected result. Even if the prior is non-informative, the obtained coverage probabilities are not too low, unlike the observed coverage probabilities of the Mat\`{e}rn field parameters which are generally low when non-informative priors are used. 

For the measurement error variance $\sigma^2_e$, the coverage probabilities appear to be smaller when there are more time points. Again, this result may seem counter-intuitive, but the explanation for this is the same for $\beta_1$. When there are more time points, there are more data points which means more information at our disposal for the estimation, and this leads to generally narrower intervals. As seen in Table \ref{tab:rmse.exposures}, for a fixed setting for the sparsity and priors specification, as we have more time points the RMSE generally decreases, and thus we also expect to have narrower coverage probabilities. Table \ref{tab:coverage.exposures} also shows that the coverage probabilities for $\sigma^2_e$ are generally high when the monitors data is sparse and when the priors specification is informative. When the data is sparse, the standard error is relatively larger because of fewer information, but since penalized complexity priors are used, the parameter estimate is very close to the true value; and this explains the very high coverage probabilities. However, if the data is not sparse, there are several data points, which potentially gives smaller standard errors and narrower credible intervals, thus the lower coverage probabilities. Nonetheless, Tables \ref{tab:bias.exposures} and \ref{tab:rmse.exposures} show that the biases and RMSEs are smaller when the monitors data is not sparse and when informative priors are used. 

For the parameters of the proxy data, the coverage probabilities of $\alpha_0$ and $\alpha_1$ are generally high for all scenarios, and are also close to the true nominal value of 95\%. For the variance parameter $\sigma_{\delta}^2$, the coverage probabilities are generally smaller when there are more time points. From Tables \ref{tab:bias.exposures} and \ref{tab:rmse.exposures}, the biases for $\sigma_{\delta}^2$ are consistently close to zero for all scenarios and the RMSEs are consistently smaller for more time points, which also means narrower credible intervals. A very narrow credible interval could be the explanation for the low coverage probabilities, but the point estimates are close to true value because of very small bias.  

\subsection{Second-stage model parameters}

Tables \ref{tab:bias.poisson} and \ref{tab:rmse.poisson} shows the biases and RMSEs of the health model parameters for all scenarios and the two methods for computing the block-level estimates of exposures.

\begin{figure}[ht]
        \centering
     	\includegraphics[trim={3cm 3cm 3cm 3cm},clip,width=1\textwidth]{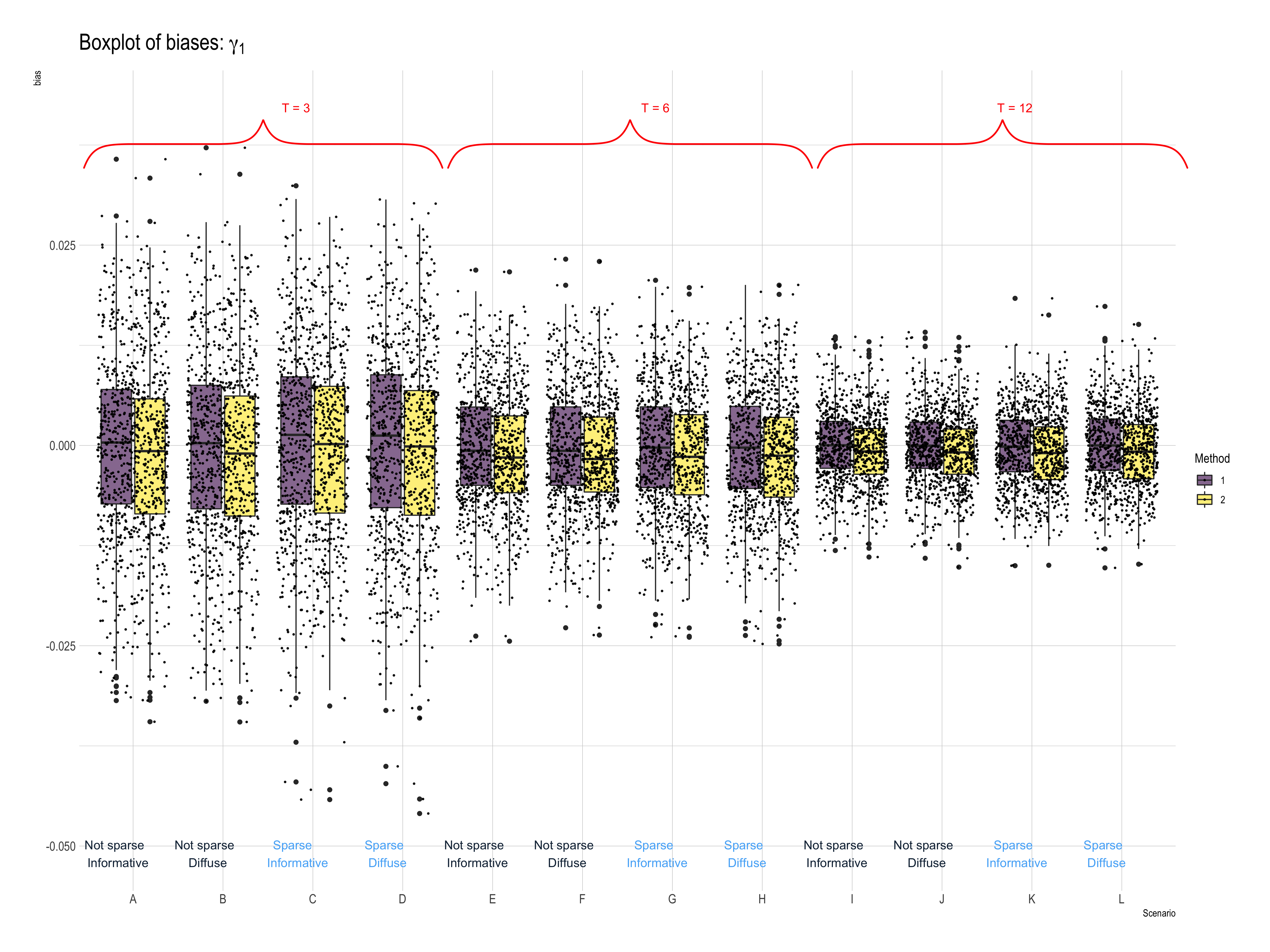}
	\caption{Plot of bias in $\gamma_1$ for all scenarios}
	\label{fig:Bias.perf.gamma1}
\end{figure}

\begin{figure}[ht]
        \centering
     	\includegraphics[trim={3cm 3cm 3cm 3cm},clip,width=1\textwidth]{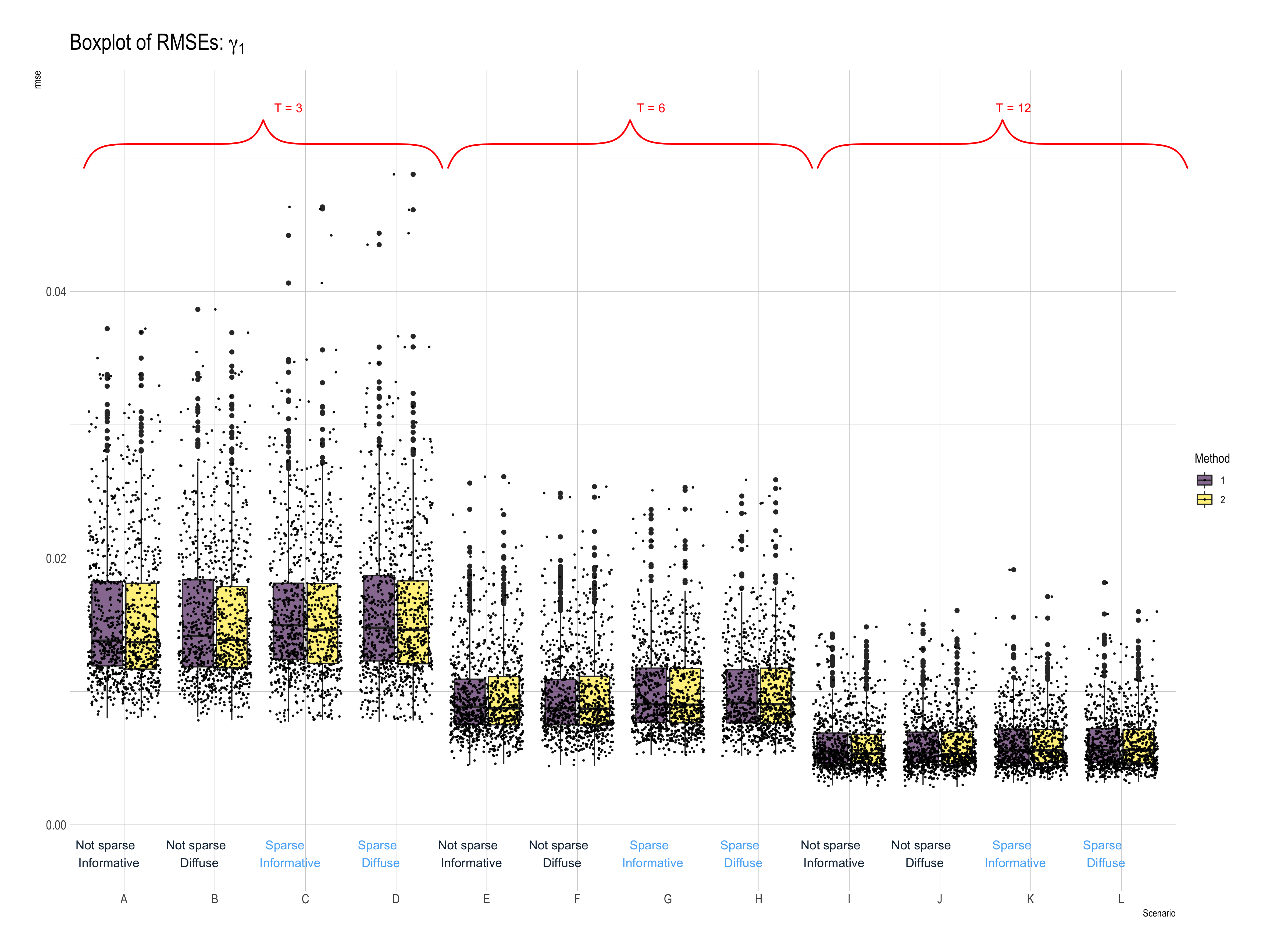}
	\caption{Plot of RMSE in $\gamma_1$ for all scenarios}
	\label{fig:RMSE.perf.gamma1}
\end{figure}

The main parameter of interest is $\gamma_1$ since it is the coefficient of the exposures in the health model. Figures \ref{fig:Bias.perf.gamma1} and \ref{fig:RMSE.perf.gamma1} show the biases and the RMSEs for $\gamma_1$ for all the scenarios. There is no striking difference in the biases and RMSEs between the two methods for computing the spatial averages of exposures. The figures show that the bias in $\gamma_1$ is close to zero for all scenarios. Also, for more time points, the RMSEs become smaller. This is expected since with more time points, there are more information available to estimate the parameter properly. Moreover, the sparsity of the monitors does not affect the quality of estimates for $\gamma_1$. This is also expected since as shown in section \ref{subsec:blocklevresults}, the obtained block-level exposures estimates are highly correlated and close to the true values of block-level exposures whether the monitors data is sparse or not. Hence, the obtained estimates for $\gamma_1$ will be similar for either case. Lastly, even with non-informative priors on $\gamma_1$, the bias and the RMSE is consistently small. Note that in the simulation study, the $\gamma_1$ parameter has a non-informative prior for all the scenarios, and so all the values shown in Figures \ref{fig:Bias.perf.gamma1} and \ref{fig:RMSE.perf.gamma1} are computed using non-informative priors for $\gamma_1$. The insights for $\gamma_1$ also holds true for the intercept $\gamma_0$, as shown in Tables \ref{tab:bias.poisson} and \ref{tab:rmse.poisson}.

For the other second-stage parameters, \ref{app:plot.bias.poisson} and \ref{app:plot.rmse.poisson} show the plots of the bias and RMSE for all scenarios. For the variance parameter of the time effect $\sigma^2_{\nu}$, the biases and RMSEs are generally smaller when there are more time points. This makes sense since $\sigma^2_{\nu}$ is a parameter of the temporal random effect, and so it is more accurately estimated when the length of the time series is longer. In addition, the prior specification also affects the precision of the estimates. If the priors used are informative, the RMSEs for $\sigma^2_{\nu}$ are generally lower compared the scenarios where the priors are non-informative. However, the difference in the RMSEs for  $\sigma^2_{\nu}$ for the informative and non-informative priors diminishes with more time points.

For the variance of the block-specific effect $\sigma^2_{\phi}$, the biases and the RMSEs are generally close to zero for all the scenarios. The number of time points does not seem to unduly affect the biases and RMSEs. This is expected since the model in the simulation study assumes that the block-specific effect is independent with time, and so the number of time points in the data does not potentially affect the accuracy and precision of the estimates of $\sigma^2_{\phi}$. In addition, this parameter does not seem to be sensitive to the prior specification and the sparsity of the monitors.  

\begin{table}[]
\resizebox{\linewidth}{!}{%
\begin{tabular}{|l|c|l|cc|cc|cc|cc|}
\hline
\multicolumn{1}{|c|}{\multirow{2}{*}{\textbf{T}}} & \multirow{2}{*}{\textbf{Sparse}} & \multicolumn{1}{c|}{\multirow{2}{*}{\textbf{Priors}}} & \multicolumn{2}{c|}{$\bm{\gamma_0}$}           & \multicolumn{2}{c|}{$\bm{\gamma_1}$}            & \multicolumn{2}{c|}{$\bm{\sigma^2_{\phi}}$}         & \multicolumn{2}{c|}{$\bm{\sigma^2_{\nu}}$}          \\ \cline{4-11} 
\multicolumn{1}{|c|}{}                            &                                  & \multicolumn{1}{c|}{}                                 & \multicolumn{1}{c|}{\textbf{M1}} & \textbf{M2} & \multicolumn{1}{c|}{\textbf{M1}} & \textbf{M2} & \multicolumn{1}{c|}{\textbf{M1}} & \textbf{M2} & \multicolumn{1}{c|}{\textbf{M1}} & \textbf{M2} \\ \hline
\multirow{4}{*}{3}                                & \multirow{2}{*}{No}              & informative                                           & \multicolumn{1}{c|}{0.0052}      & 0.0050      & \multicolumn{1}{c|}{0.0000}      & -0.0012     & \multicolumn{1}{c|}{0.0001}      & 0.0000      & \multicolumn{1}{c|}{-0.0039}     & -0.0039     \\ \cline{3-11} 
                                                  &                                  & non-informative                                       & \multicolumn{1}{c|}{0.0052}      & 0.0053      & \multicolumn{1}{c|}{0.0000}      & -0.0012     & \multicolumn{1}{c|}{0.0001}      & 0.0000      & \multicolumn{1}{c|}{-0.0018}     & -0.0018     \\ \cline{2-11} 
                                                  & \multirow{2}{*}{Yes}             & informative                                           & \multicolumn{1}{c|}{-0.0009}     & -0.0010     & \multicolumn{1}{c|}{0.0009}      & -0.0004     & \multicolumn{1}{c|}{0.0004}      & 0.0003      & \multicolumn{1}{c|}{-0.0047}     & -0.0047     \\ \cline{3-11} 
                                                  &                                  & non-informative                                       & \multicolumn{1}{c|}{-0.0010}     & -0.0013     & \multicolumn{1}{c|}{0.0005}      & -0.0008     & \multicolumn{1}{c|}{0.0004}      & 0.0003      & \multicolumn{1}{c|}{-0.0031}     & -0.0031     \\ \hline
\multirow{4}{*}{6}                                & \multirow{2}{*}{No}              & informative                                           & \multicolumn{1}{c|}{0.0009}      & 0.0006      & \multicolumn{1}{c|}{0.0000}      & -0.0011     & \multicolumn{1}{c|}{-0.0002}     & -0.0003     & \multicolumn{1}{c|}{-0.0018}     & -0.0018     \\ \cline{3-11} 
                                                  &                                  & non-informative                                       & \multicolumn{1}{c|}{0.0009}      & 0.0006      & \multicolumn{1}{c|}{-0.0001}     & -0.0012     & \multicolumn{1}{c|}{-0.0002}     & -0.0003     & \multicolumn{1}{c|}{-0.0007}     & -0.0007     \\ \cline{2-11} 
                                                  & \multirow{2}{*}{Yes}             & informative                                           & \multicolumn{1}{c|}{0.0004}      & 0.0002      & \multicolumn{1}{c|}{-0.0002}     & -0.0013     & \multicolumn{1}{c|}{0.0002}      & 0.0001      & \multicolumn{1}{c|}{-0.0017}     & -0.0017     \\ \cline{3-11} 
                                                  &                                  & non-informative                                       & \multicolumn{1}{c|}{0.0003}      & 0.0002      & \multicolumn{1}{c|}{-0.0003}     & -0.0014     & \multicolumn{1}{c|}{0.0002}      & 0.0002      & \multicolumn{1}{c|}{-0.0007}     & -0.0007     \\ \hline
\multirow{4}{*}{12}                               & \multirow{2}{*}{No}              & informative                                           & \multicolumn{1}{c|}{-0.0018}     & -0.0019     & \multicolumn{1}{c|}{0.0001}      & -0.0007     & \multicolumn{1}{c|}{0.0001}      & 0.0000      & \multicolumn{1}{c|}{-0.0004}     & -0.0004     \\ \cline{3-11} 
                                                  &                                  & non-informative                                       & \multicolumn{1}{c|}{-0.0018}     & -0.0019     & \multicolumn{1}{c|}{0.0001}      & -0.0008     & \multicolumn{1}{c|}{0.0001}      & 0.0000      & \multicolumn{1}{c|}{0.0001}      & 0.0001      \\ \cline{2-11} 
                                                  & \multirow{2}{*}{Yes}             & informative                                           & \multicolumn{1}{c|}{0.0019}      & 0.0016      & \multicolumn{1}{c|}{-0.0001}     & -0.0010     & \multicolumn{1}{c|}{0.0001}      & 0.0000      & \multicolumn{1}{c|}{-0.0002}     & -0.0002     \\ \cline{3-11} 
                                                  &                                  & non-informative                                       & \multicolumn{1}{c|}{0.0019}      & 0.0016      & \multicolumn{1}{c|}{0.0002}      & -0.0007     & \multicolumn{1}{c|}{0.0001}      & 0.0001      & \multicolumn{1}{c|}{0.0003}      & 0.0003      \\ \hline
\end{tabular}}
\caption{\label{tab:bias.poisson}Bias of health model parameters for all scenarios}
\end{table}

\begin{table}[]
\resizebox{\linewidth}{!}{%
\begin{tabular}{|l|c|l|ll|ll|ll|ll|}
\hline
\multicolumn{1}{|c|}{\multirow{2}{*}{\textbf{T}}} & \multirow{2}{*}{\textbf{Sparse}} & \multicolumn{1}{c|}{\multirow{2}{*}{\textbf{Priors}}} & \multicolumn{2}{c|}{$\bm{\gamma_0}$}                                & \multicolumn{2}{c|}{$\bm{\gamma_1}$}                                 & \multicolumn{2}{c|}{$\bm{\sigma_{\phi}^2}$}                              & \multicolumn{2}{c|}{$\bm{\sigma_{\nu}^2}$}                               \\ \cline{4-11} 
\multicolumn{1}{|c|}{}                            &                                  & \multicolumn{1}{c|}{}                                 & \multicolumn{1}{c|}{\textbf{M1}} & \multicolumn{1}{c|}{\textbf{M2}} & \multicolumn{1}{c|}{\textbf{M1}} & \multicolumn{1}{c|}{\textbf{M2}} & \multicolumn{1}{c|}{\textbf{M1}} & \multicolumn{1}{c|}{\textbf{M2}} & \multicolumn{1}{c|}{\textbf{M1}} & \multicolumn{1}{c|}{\textbf{M2}} \\ \hline
\multirow{4}{*}{3}                                & \multirow{2}{*}{No}              & informative                                           & \multicolumn{1}{l|}{0.1053}      & 0.1051                           & \multicolumn{1}{l|}{0.0156}      & 0.0154                           & \multicolumn{1}{l|}{0.0051}      & 0.0051                           & \multicolumn{1}{l|}{0.0168}      & 0.0168                           \\ \cline{3-11} 
                                                  &                                  & non-informative                                       & \multicolumn{1}{l|}{0.1097}      & 0.1099                           & \multicolumn{1}{l|}{0.0156}      & 0.0155                           & \multicolumn{1}{l|}{0.0052}      & 0.0052                           & \multicolumn{1}{l|}{0.0247}      & 0.0246                           \\ \cline{2-11} 
                                                  & \multirow{2}{*}{Yes}             & informative                                           & \multicolumn{1}{l|}{0.1033}      & 0.1036                           & \multicolumn{1}{l|}{0.0161}      & 0.0158                           & \multicolumn{1}{l|}{0.0051}      & 0.0051                           & \multicolumn{1}{l|}{0.0165}      & 0.0164                           \\ \cline{3-11} 
                                                  &                                  & non-informative                                       & \multicolumn{1}{l|}{0.1068}      & 0.1066                           & \multicolumn{1}{l|}{0.0162}      & 0.0159                           & \multicolumn{1}{l|}{0.0052}      & 0.0052                           & \multicolumn{1}{l|}{0.0234}      & 0.0235                           \\ \hline
\multirow{4}{*}{6}                                & \multirow{2}{*}{No}              & informative                                           & \multicolumn{1}{l|}{0.0776}      & 0.0777                           & \multicolumn{1}{l|}{0.0097}      & 0.0096                           & \multicolumn{1}{l|}{0.0046}      & 0.0046                           & \multicolumn{1}{l|}{0.0135}      & 0.0135                           \\ \cline{3-11} 
                                                  &                                  & non-informative                                       & \multicolumn{1}{l|}{0.0789}      & 0.0790                           & \multicolumn{1}{l|}{0.0097}      & 0.0096                           & \multicolumn{1}{l|}{0.0047}      & 0.0047                           & \multicolumn{1}{l|}{0.0167}      & 0.0167                           \\ \cline{2-11} 
                                                  & \multirow{2}{*}{Yes}             & informative                                           & \multicolumn{1}{l|}{0.0783}      & 0.0784                           & \multicolumn{1}{l|}{0.0100}      & 0.0099                           & \multicolumn{1}{l|}{0.0046}      & 0.0046                           & \multicolumn{1}{l|}{0.0135}      & 0.0135                           \\ \cline{3-11} 
                                                  &                                  & non-informative                                       & \multicolumn{1}{l|}{0.0798}      & 0.0797                           & \multicolumn{1}{l|}{0.0101}      & 0.0100                           & \multicolumn{1}{l|}{0.0047}      & 0.0047                           & \multicolumn{1}{l|}{0.0166}      & 0.0166                           \\ \hline
\multirow{4}{*}{12}                               & \multirow{2}{*}{No}              & informative                                           & \multicolumn{1}{l|}{0.0582}      & 0.0584                           & \multicolumn{1}{l|}{0.0060}      & 0.0060                           & \multicolumn{1}{l|}{0.0042}      & 0.0042                           & \multicolumn{1}{l|}{0.0105}      & 0.0105                           \\ \cline{3-11} 
                                                  &                                  & non-informative                                       & \multicolumn{1}{l|}{0.0585}      & 0.0585                           & \multicolumn{1}{l|}{0.0060}      & 0.0060                           & \multicolumn{1}{l|}{0.0043}      & 0.0043                           & \multicolumn{1}{l|}{0.0117}      & 0.0118                           \\ \cline{2-11} 
                                                  & \multirow{2}{*}{Yes}             & informative                                           & \multicolumn{1}{l|}{0.0586}      & 0.0586                           & \multicolumn{1}{l|}{0.0062}      & 0.0062                           & \multicolumn{1}{l|}{0.0043}      & 0.0043                           & \multicolumn{1}{l|}{0.0105}      & 0.0106                           \\ \cline{3-11} 
                                                  &                                  & non-informative                                       & \multicolumn{1}{l|}{0.0587}      & 0.0587                           & \multicolumn{1}{l|}{0.0062}      & 0.0062                           & \multicolumn{1}{l|}{0.0044}      & 0.0044                           & \multicolumn{1}{l|}{0.0118}      & 0.0118                           \\ \hline
\end{tabular}}
\caption{\label{tab:rmse.poisson}RMSEs of health model parameters for all scenarios}
\end{table}

Table \ref{tab:coverage.poisson} shows the coverage probabilities for the health model parameters. For $\gamma_1$, all the coverage probabilities are very close to the nominal value of 95\%. There is no difference between the two methods of computing spatial averages. For the intercept $\gamma_0$, the coverage probabilities are also reasonably close to the nominal value. The coverage probabilities for scenarios with fewer time points is smaller compared to the scenarios with more time points. The coverage probabilities for the variance parameter of the time random effect $\sigma^2_{\nu}$ is higher when penalized complexity priors are used or when there are more time points. Finally, for the variance of the block-specific random effect $\sigma^2_{\phi}$, the coverage probabilities are consistently and high and close to the nominal value for all scenarios.

\begin{table}[]
\resizebox{\linewidth}{!}{%
\begin{tabular}{|l|c|l|cc|cc|cc|cc|}
\hline
\multicolumn{1}{|c|}{\multirow{2}{*}{\textbf{T}}} & \multirow{2}{*}{\textbf{Sparse}} & \multicolumn{1}{c|}{\multirow{2}{*}{\textbf{Priors}}} & \multicolumn{2}{c|}{$\bm{\gamma_0}$}           & \multicolumn{2}{c|}{$\bm{\gamma_1}$}            & \multicolumn{2}{c|}{$\bm{\sigma_{\phi}^2}$}         & \multicolumn{2}{c|}{$\bm{\sigma_{\nu}^2}$}          \\ \cline{4-11} 
\multicolumn{1}{|c|}{}                            &                                  & \multicolumn{1}{c|}{}                                 & \multicolumn{1}{c|}{\textbf{M1}} & \textbf{M2} & \multicolumn{1}{c|}{\textbf{M1}} & \textbf{M2} & \multicolumn{1}{c|}{\textbf{M1}} & \textbf{M2} & \multicolumn{1}{c|}{\textbf{M1}} & \textbf{M2} \\ \hline
\multirow{4}{*}{3}                                & \multirow{2}{*}{No}              & informative                                           & \multicolumn{1}{c|}{85.4}        & 85.2        & \multicolumn{1}{c|}{94.2}        & 93          & \multicolumn{1}{c|}{95.2}        & 93.8        & \multicolumn{1}{c|}{88.4}        & 88.6        \\ \cline{3-11} 
                                                  &                                  & non-informative                                       & \multicolumn{1}{c|}{81.8}        & 81.2        & \multicolumn{1}{c|}{93.6}        & 93.4        & \multicolumn{1}{c|}{94.8}        & 93.4        & \multicolumn{1}{c|}{71.6}        & 71.6        \\ \cline{2-11} 
                                                  & \multirow{2}{*}{Yes}             & informative                                           & \multicolumn{1}{c|}{86.4}        & 85.8        & \multicolumn{1}{c|}{94}          & 95          & \multicolumn{1}{c|}{95}          & 95.4        & \multicolumn{1}{c|}{88}          & 88          \\ \cline{3-11} 
                                                  &                                  & non-informative                                       & \multicolumn{1}{c|}{83}          & 82.6        & \multicolumn{1}{c|}{93.2}        & 93.8        & \multicolumn{1}{c|}{94.2}        & 94.6        & \multicolumn{1}{c|}{71.2}        & 71          \\ \hline
\multirow{4}{*}{6}                                & \multirow{2}{*}{No}              & informative                                           & \multicolumn{1}{c|}{91.8}        & 91.6        & \multicolumn{1}{c|}{93.6}        & 94          & \multicolumn{1}{c|}{94.2}        & 93          & \multicolumn{1}{c|}{89.8}        & 89.4        \\ \cline{3-11} 
                                                  &                                  & non-informative                                       & \multicolumn{1}{c|}{91.4}        & 91.2        & \multicolumn{1}{c|}{94}          & 94          & \multicolumn{1}{c|}{94}          & 92.6        & \multicolumn{1}{c|}{84.2}        & 83.4        \\ \cline{2-11} 
                                                  & \multirow{2}{*}{Yes}             & informative                                           & \multicolumn{1}{c|}{91}          & 90.6        & \multicolumn{1}{c|}{94.4}        & 94          & \multicolumn{1}{c|}{94.2}        & 94          & \multicolumn{1}{c|}{89.6}        & 90.6        \\ \cline{3-11} 
                                                  &                                  & non-informative                                       & \multicolumn{1}{c|}{90}          & 90.4        & \multicolumn{1}{c|}{94.2}        & 94          & \multicolumn{1}{c|}{94.2}        & 93.6        & \multicolumn{1}{c|}{86}          & 85.4        \\ \hline
\multirow{4}{*}{12}                               & \multirow{2}{*}{No}              & informative                                           & \multicolumn{1}{c|}{92.6}        & 93.2        & \multicolumn{1}{c|}{94}          & 93.8        & \multicolumn{1}{c|}{94.2}        & 95.2        & \multicolumn{1}{c|}{92.4}        & 93          \\ \cline{3-11} 
                                                  &                                  & non-informative                                       & \multicolumn{1}{c|}{93.4}        & 92.8        & \multicolumn{1}{c|}{94}          & 94.4        & \multicolumn{1}{c|}{94.2}        & 94.6        & \multicolumn{1}{c|}{90}          & 90          \\ \cline{2-11} 
                                                  & \multirow{2}{*}{Yes}             & informative                                           & \multicolumn{1}{c|}{94.8}        & 94.6        & \multicolumn{1}{c|}{95.4}        & 93.8        & \multicolumn{1}{c|}{93.6}        & 93.4        & \multicolumn{1}{c|}{92.8}        & 92          \\ \cline{3-11} 
                                                  &                                  & non-informative                                       & \multicolumn{1}{c|}{93.6}        & 94.2        & \multicolumn{1}{c|}{93.4}        & 93.6        & \multicolumn{1}{c|}{93.2}        & 93.4        & \multicolumn{1}{c|}{89.8}        & 89.2        \\ \hline
\end{tabular}}
\caption{\label{tab:coverage.poisson}Coverage probabilities of health model parameters for all scenarios}
\end{table}

\subsection{Block-level exposure estimates} \label{subsec:blocklevresults}

\begin{figure}[ht]
        \centering
     	\includegraphics[trim={0 0 0 3cm},clip,width=\textwidth]{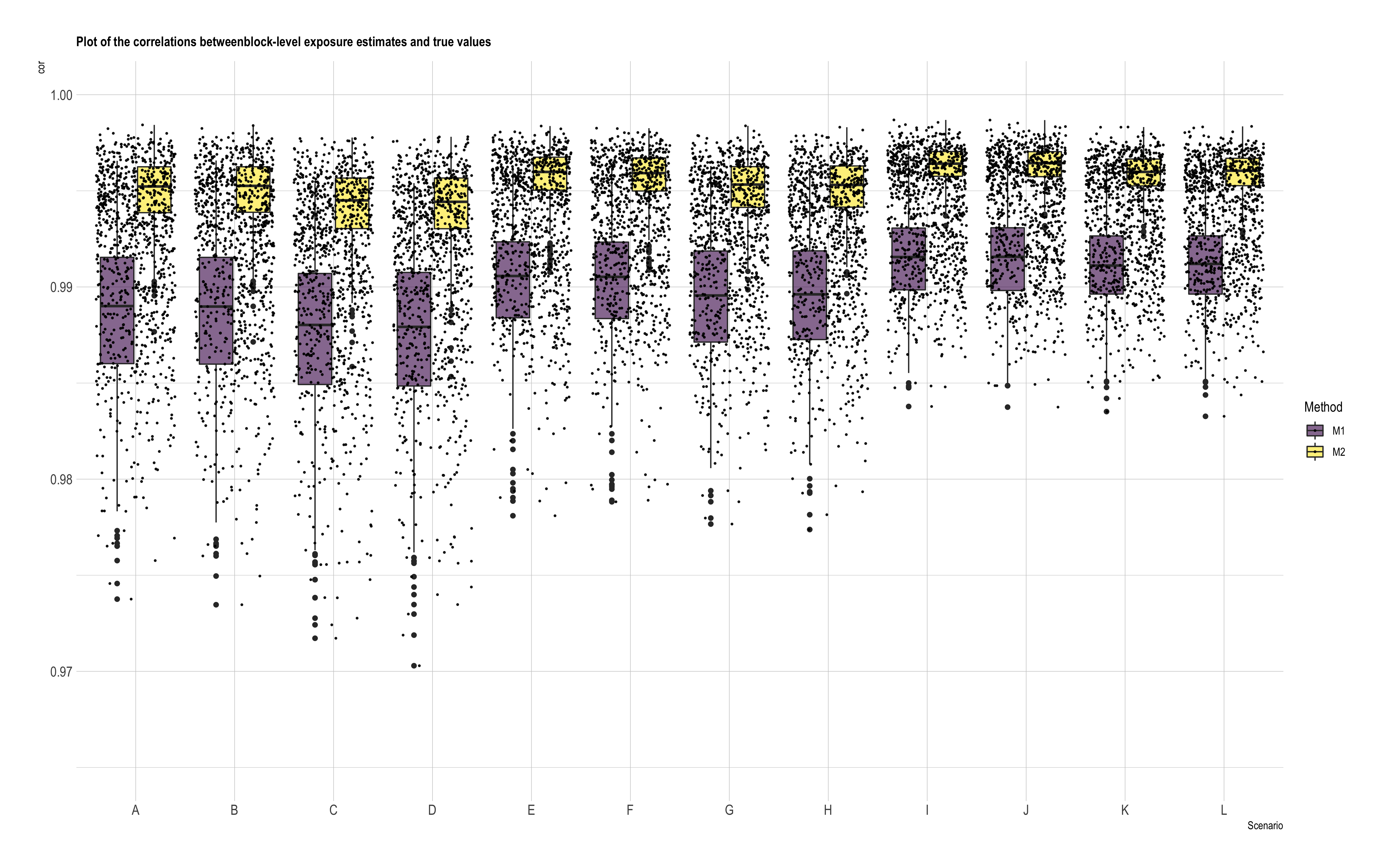}
	\caption{Correlations between block-level exposure estimates and true values}
	\label{fig:blocks.corrs}
\end{figure}

Figure \ref{fig:blocks.corrs} shows the correlations between the block-level exposure estimates and the corresponding true values for all scenarios. All the correlations ranges from around 0.97 to some value close to 1.0, but using method 2 for computing spatial averages generally gave higher correlations than method 1 for all scenarios. Also, the correlations are generally higher when there are more time points, which is true for both methods. 

\begin{figure}[ht]
        \centering
     	\includegraphics[trim={0 0 0 3cm},clip,width=\textwidth]{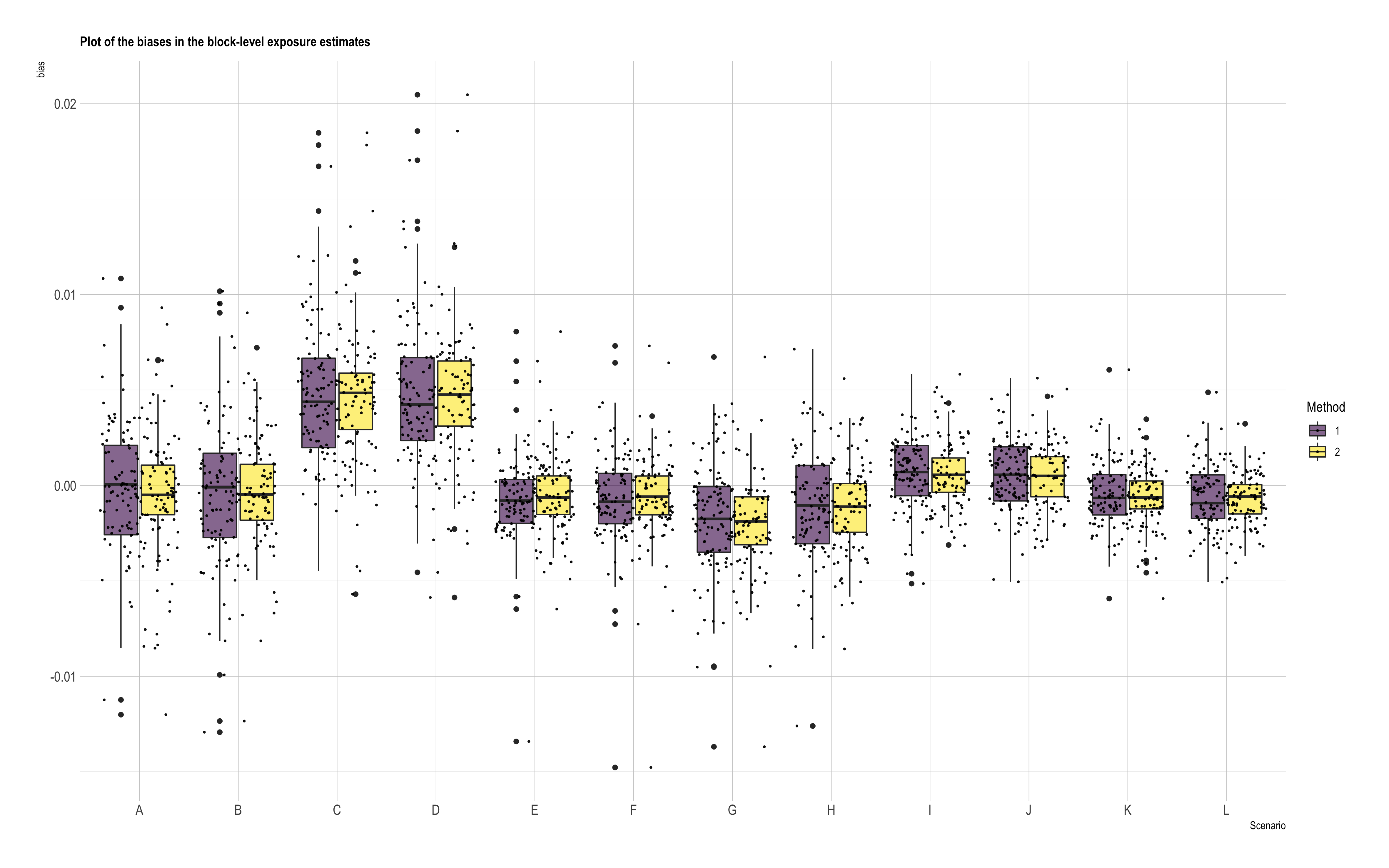}
	\caption{Biases in block-level exposure estimates for all scenarios}
	\label{fig:blocks.bias}
\end{figure}

Figure \ref{fig:blocks.bias} shows the biases in the block-level exposure estimates for all the scenarios. Each point in the plot corresponds to an area or block in the simulation study region. The bias for each block is computed as the average of the difference between the true block exposure value and its estimated value across all the 500 replicates. Since there are $T$ block-level estimates of exposures for a block in a specific simulated data, the final values shown in Figure \ref{fig:blocks.bias} are the average of the biases across time for each block. The average bias for all 98 areas are generally close to zero. Moreover, the spread in the biases are wider for method 1 than method 2. Also, when the data on the monitors is sparse and the priors are non-informative, the biases are generally larger, especially for scenarios with fewer time points.

\begin{figure}[ht]
        \centering
     	\includegraphics[trim={0 0 0 3cm},clip,width=\textwidth]{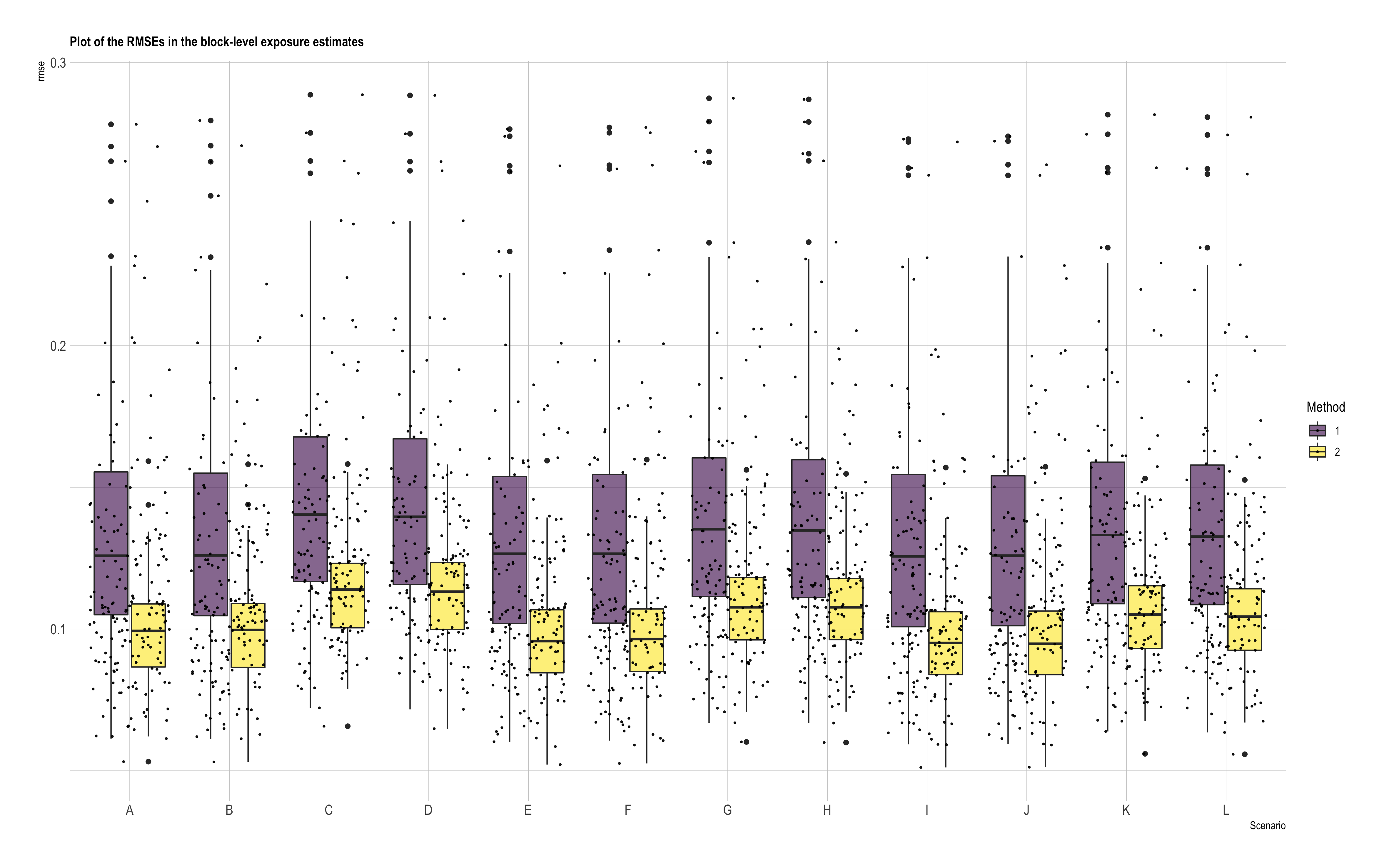}
	\caption{RMSEs in block-level exposure estimates for all scenarios}
	\label{fig:blocks.rmse}
\end{figure}

The observed pattern in Figure \ref{fig:blocks.bias}, in which the spread of the bias is higher when using method 1 than method 2, is consistent with the RMSEs shown in Figure \ref{fig:blocks.rmse}. As in Figure \ref{fig:blocks.bias}, each point in Figure \ref{fig:blocks.rmse} corresponds to the average RMSE of an area, averaged over the $T$ time points and computed from the 500 replicates. There are a couple of areas with very high RMSEs when using method 1 for computing block-level spatial average, and this is consistent for all the scenarios. The average of the RMSEs from using method 1 is higher compared to method 2 for all the scenarios. 

\section{Conclusions}

This paper presents a two-stage spatio-temporal model for the epidemiological problem of estimating the effect of exposures on health outcomes where the data have different spatial supports. The framework of the first-stage model is based on the Bayesian melding model for which a common latent field is assumed for the geostatistical data on the monitors and the high-resolution proxy data, for which both are error-prone realizations of the latent field.  However, a non-fully Bayesian melding approach is adopted in this work since the proxy data is treated as a geostatistical data on the centroids of the grids. A couple of reasons for this simplification is the assumption that the resolution of the covariate data is not finer than that of the proxy data, that the resolution of the proxy data is fine enough that it can be treated as geostatistical at the centroids, and also to simplify the computation. Nonetheless, the proposed model incorporates a non-spatially varying additive and multiplicative bias, and an additional additive noise in the proxy data with the assumption that it is noisier and less correlated with the true latent field compared to the measurements from the network of monitoring stations. The SPDE approach is used to model the spatio-temporal structure of the model in order to speed up computation and spatial interpolation. The second stage fits a GLMM using the spatial averages of the latent field, and additional spatial and temporal random effects. Both the first-stage and second-stage models, being latent Gaussian, are fitted using the integrated nested Laplace approximation. In estimating the block-level exposure estimates, two methods proposed in \cite{inlaspde} were used. These block-level estimates of exposures are then inputted in the second stage health model. In order to account for the uncertainty in the block-level estimates when fitting the second-stage model, samples from the marginal posterior distribution of the latent field at the prediction grid are generated. For each set of simulated values, block-level estimates of exposures are computed using the two methods and are then used to fit the second-stage model. This is done several times and then all results are combined and used to approximate the posterior distribution of the second-stage model parameters. 

A simulation study was carried out to assess the performance of the proposed method under different scenarios. Considered in the study are the following settings: the sparsity of the data on the monitors, the number of time points, and the prior specification. It is common to work with sparse data on the monitors so it is interesting to look at the effect of the sparsity on the quality of the parameter estimates. Also, it is important in Bayesian analysis to assess the sensitivity of the results to the priors, more so in the context of a complex spatio-temporal process with several model parameters that need to be estimated. 

All first-stage model parameters have generally small biases but there is great difficulty in estimating the latent field parameters, particularly the spatial variance, the range parameter, and the autoregressive parameter especially if non-informative priors are used. If informative priors are used, the bias and RSMEs are very small and the coverage probabilities are very high even if the data on the monitors is sparse. 

For the main parameter of interest, $\gamma_1$, the method provides very good estimates across all scenarios considered in the simulation study. There is no difference between the two methods of computing the spatial averages in terms of the bias, RMSEs, and coverage probabilities. Even with non-informative prior on $\gamma_1$ and sparse data on the monitors, the estimates for $\gamma_1$ are close to the true value. Finally, with more time points, the RMSEs were observed to decrease. 

The simulation study done also showed that the sparsity of the data on the monitors can potentially affect the quality of the parameter estimates. When the data on the monitors is sparse, the RMSE of the covariate effect in the latent field is large. This is also true for the measurement error variance in the monitors but the use of informative priors can be helpful to accurately estimate the parameter. It makes sense for these two parameters to be seriously affected by the sparsity of the monitors data since these two parameters are directly linked to the observed values at the monitors; hence, the more monitors in the study map, the more information at our disposal to accurately estimate these two parameters. The RMSEs of the Mat\`{e}rn covariance parameters and the bias parameters of the proxy data are also generally higher when the monitors data is sparse. The parameters in the second-stage model seem to be not affected by the sparsity of the data on the monitors since the proposed method, even with sparse data, was able to estimate well the latent field and that the block-level predicted values are also close to the true value, at least for most of the areas, as shown in the correlations, biases, and RMSEs from the simulation results.  However, in practice, this will only be true for as long as the block-level estimates of exposures are close to the unknown block averages of the latent continuous process.

The use of informative priors gave better parameter estimates especially for the latent field parameters which are typically the most difficult parameters to be estimated. When informative priors are used, the estimates of the spatial field variance and the range parameter have lower bias and RMSE. The autoregressive parameter of the latent field also benefits with the use of informative priors, giving smaller errors and higher coverage probabilities. In addition, the measurement error variance at the monitors also have lower RMSEs with informative priors especially for the case when the data is sparse. The rest of the parameters are not too sensitive to the prior specification which includes the variance of the block-specific random effect in the second-stage model.

The number of time points can also potentially affect the quality of the estimates. As already mentioned, if there are more time points, the RMSEs of the fixed effects in the second-stage model are smaller. This is also true for the variance of the time effect in the second-stage model. The bias parameters of the proxy data in the first-stage model also have better estimates with more time points.

The method for computing block-level exposures estimates does not show to have an impact on the parameter estimates. The biases in the block-level estimates for both methods are on the average close to zero, but the first method seem to give higher biases for certain blocks and also higher RMSEs overall. Also, the second method gave higher correlations between the true block-level exposure values and the estimated values. Nonetheless, both methods gave fairly highly correlations which are at least 0.97. And as far as the parameter estimation is concerned, either of the two methods should work very well. 

The specification of the spatial and temporal effects and their possible interactions in the second-stage model can be easily extended to more complex models such as the use of conditional autoregressive models and spatio-temporal interaction effects. A possible drawback of the proposed method is that it becomes computationally expensive with a long time series data since it will significantly increase the number of unobserved values of the latent field that need to be estimated. This would primarily depend on the resolution of the proxy data and also the size of the study region. 

The use of a fully Bayesian melding approach would all the more increase the computational costs because of the enlarged latent field vector and the extended latent field to accommodate the `copy' approach when fitting the first-stage joint model. Even with the use of INLA which is a fast approximation-based Bayesian inference method and the use of the SPDE approach which induces a Markov structure in the continuous spatial process, the number of parameters and the unobserved values of the latent field as well as the values in the prediction grid which is used to compute the block-level estimates altogether is high-dimensional which will certainly amplify the computation effort required. Under a full Bayesian melding model, it is required to have covariate information on $\bm{z}(\bm{s},t)$ at a resolution which is finer than the resolution of the proxy data. Also, using the data augmentation approach, this would require an enlarged latent vector, say $\bm{x}_{t,P^*}$ from $\bm{x}_{t,P}$ in equation \eqref{pseudozeroes}, with dimension equal to the number number of data points in the covariate $\bm{z}(\bm{s},t)$, which should be greater than the number of grid points of the proxy data. This also implies an enlarged extended field, say $\bm{\chi}_t= \begin{pmatrix} \bm{x}^{\intercal}_{t,M} & \bm{x}^{\intercal}_{t,P^*} & \bm{x}^{*\intercal}_{t,M} & \bm{x}^{*\intercal}_{t,P^*} \end{pmatrix}^\intercal, t=1,\ldots,T$, where $\bm{x}^{*\intercal}_{t,M}$ is a copy of $\bm{x}^{\intercal}_{t,M}$ and $\bm{x}^{*\intercal}_{t,P^*}$ is a copy of $\bm{x}^{\intercal}_{t,P^*}$. Using the fact that
\begin{align} x(B,t) &=\dfrac{1}{|B|}\int_{\forall \bm{s} \in B}  \bigg( \alpha_0 + \alpha_1  x(\bm{s},t)+ \delta(\bm{s},t) \bigg) d\bm{s},
			  \end{align}
the joint model is now given as follows:
\begin{align}
	\bm{w}_t &= \bm{x}^*_{t,M} + \bm{e}_t, \;\;\; \bm{e}_t\sim N(\textbf{0},\sigma^2_{e}\mathbb{I}_M), \;\; t=1,\ldots, T \\\
x(B,t) &= \alpha_0 + \dfrac{1}{|B|} \int_{|B|} \alpha_1x^*(\bm{s},t) +  \dfrac{1}{|B|} \int_{|B|} \delta(\bm{s},t), \;\; t=1,\ldots, T \label{eq:fullmelding} \\
\bm{0}_t &= -\begin{pmatrix} \bm{x}_{t,M} \\ \bm{x}_{t,P^*} \end{pmatrix} + \beta_{0}\mathbf{1}_G + \beta_{1}\bm{z}_t+ \bm{B} \bm{\xi}_t^D, \;\; t=1,\ldots, T \\				
\bm{\xi}_t^D &= \varsigma \bm{\xi}_{t-1}^D + \bm{\omega}_t^D, \;\;\; \bm{\omega}_t^D \sim N(\bm{0},\bm{Q}_s^{-1}), \;\;\; t=1,\ldots,T
\end{align}
where $x^*(\bm{s},t)$ is a `copy' of an element in the latent vector $\bm{x}_{t,P^*}$.
In equation \eqref{eq:fullmelding}, $\dfrac{1}{|B|} \bigintsss_{|B|} x^*(\bm{s},t)$ is approximated as
\begin{align}
    \dfrac{1}{|B|} \int_{|B|} x^*(\bm{s},t) \approx \dfrac{1}{\bm{c}^\intercal \bm{1}} \bm{c}^\intercal \bm{x}^*_{t,P^*},
\end{align}
where $\bm{c}$ is a mapping vector with elements `1' or `0' which computes the sum of the elements of $\bm{x}^*_{t,P^*}$ corresponding to the points inside the grid $B$. The approach presented in section \ref{subsec:proposed.stage1.model} needs to be further extended in order to fit the full Bayesian melding joint model. The copy approach to fit equation \eqref{eq:fullmelding} could be avoided by using the SPDE representation for $x(\bm{s},t)$. However, this leads to identifiability issues for $\alpha_0$ and $\alpha_1$, and possibly for $\sigma^2_{\delta}$ since the form of $x(B,t)$ becomes
\begin{align} x(B,t) &=\dfrac{1}{|B|}\int_{B}  \bigg( \alpha_0 + \alpha_1
x(\bm{s},t)+ \delta(\bm{s},t) \bigg) d\bm{s}\\
&= \alpha_0 + \alpha_1 \dfrac{1}{|B|} \int_B \bigg( \beta_0 + \beta_1 z(\bm{s},t) + \xi(\bm{s},t)\bigg) d\bm{s} + \dfrac{1}{|B|}\int_B \delta(\bm{s},t) \\
&= \alpha_0 + \alpha_1 \beta_0 + \alpha_1\beta_1 \Bar{z}(B,t) + \alpha_1 \dfrac{1}{|B|}\int_B \xi(\bm{s},t)d\bm{s} + \dfrac{1}{|B|}\int_B \delta(\bm{s},t),
			  \end{align}
where $\dfrac{1}{|B|}\bigintsss_B \xi(\bm{s},t)d\bm{s}$ is given the SPDE representation. The use of a full Bayesian melding model is certainly a problem that should be investigated next. These are promising for future research and will be extensions of the current model assumptions and the proposed method in this paper. 

Future work may also look into how to incorporate more than one pollutant in a single spatio-temporal model. The proxy data and the locations of the monitors for each pollutant need not be aligned in space, but this will most likely require the use of two different but possibly correlated latent processes. Also, the current model assumes that the latent spatio-temporal process is constant through time. But the spatio-temporal process could evolve at some point, i.e., the Mat\`{e}rn field parameters could change, or the mean structure of the model could evolve as well. This is now a change-point detection problem which is also a promising future work since only very few work has been done for change-point detection in spatio-temporal processes. Finally, approaches to improve computational speed and efficiency especially for long time series is a very interesting direction, and also very useful for many practitioners who work with big spatio-temporal data.

\bibliography{mybibliography2}

\begin{thebibliography}{10}
\expandafter\ifx\csname url\endcsname\relax
  \def\url#1{\texttt{#1}}\fi
\expandafter\ifx\csname urlprefix\endcsname\relax\def\urlprefix{URL }\fi
\expandafter\ifx\csname href\endcsname\relax
  \def\href#1#2{#2} \def\path#1{#1}\fi

\bibitem{lawson2016handbook}
A.~B. Lawson, S.~Banerjee, R.~P. Haining, M.~D. Ugarte, Handbook of spatial
  epidemiology, CRC Press, 2016.

\bibitem{inlaspde}
M.~Cameletti, V.~G{\'o}mez-Rubio, M.~Blangiardo, Bayesian modelling for
  spatially misaligned health and air pollution data through the inla-spde
  approach, Spatial Statistics 31 (2019) 100353.

\bibitem{BLANGIARDO20161}
M.~Blangiardo, F.~Finazzi, M.~Cameletti, Two-stage bayesian model to evaluate
  the effect of air pollution on chronic respiratory diseases using drug
  prescriptions, Spatial and spatio-temporal epidemiology 18 (2016) 1--12.

\bibitem{LeeMukhopadhyayetal}
D.~Lee, S.~Mukhopadhyay, A.~Rushworth, S.~K. Sahu, A rigorous statistical
  framework for spatio-temporal pollution prediction and estimation of its
  long-term impact on health, Biostatistics 18~(2) (2017) 370--385.

\bibitem{BRUNO2016276}
F.~Bruno, M.~Cameletti, M.~Franco-Villoria, F.~Greco, R.~Ignaccolo,
  L.~Ippoliti, P.~Valentini, M.~Ventrucci, A survey on ecological regression
  for health hazard associated with air pollution, Spatial statistics 18 (2016)
  276--299.

\bibitem{diggle}
P.~J. Diggle, Statistical analysis of spatial and spatio-temporal point
  patterns, CRC press, 2013.

\bibitem{molitor}
J.~Molitor, N.-T. Molitor, M.~Jerrett, R.~McConnell, J.~Gauderman, K.~Berhane,
  D.~Thomas, Bayesian modeling of air pollution health effects with missing
  exposure data, American journal of epidemiology 164~(1) (2006) 69--76.

\bibitem{gotwayoung}
C.~A. Gotway, L.~J. Young, Combining incompatible spatial data, Journal of the
  American Statistical Association 97~(458) (2002) 632--648.

\bibitem{Leeetal}
A.~Lee, A.~Szpiro, S.~Kim, L.~Sheppard, Impact of preferential sampling on
  exposure prediction and health effect inference in the context of air
  pollution epidemiology, Environmetrics 26~(4) (2015) 255--267.

\bibitem{Kralletal}
J.~R. Krall, H.~H. Chang, S.~E. Sarnat, R.~D. Peng, L.~A. Waller, Current
  methods and challenges for epidemiological studies of the associations
  between chemical constituents of particulate matter and health, Current
  environmental health reports 2~(4) (2015) 388--398.

\bibitem{berrocal}
V.~J. Berrocal, A.~E. Gelfand, D.~M. Holland, Space-time data fusion under
  error in computer model output: An application to modeling air quality,
  Biometrics 68~(3) (2012) 837--848.

\bibitem{szpiropaciorek}
A.~A. Szpiro, C.~J. Paciorek, Measurement error in two-stage analyses, with
  application to air pollution epidemiology, Environmetrics 24~(8) (2013)
  501--517.

\bibitem{bergenszpiro}
S.~Bergen, A.~A. Szpiro, Mitigating the impact of measurement error when using
  penalized regression to model exposure in two-stage air pollution
  epidemiology studies, Environmental and ecological statistics 22~(3) (2015)
  601--631.

\bibitem{fuentes2005model}
M.~Fuentes, A.~E. Raftery, Model evaluation and spatial interpolation by
  bayesian combination of observations with outputs from numerical models,
  Biometrics 61~(1) (2005) 36--45.

\bibitem{merrorspatmis}
A.~Gryparis, C.~J. Paciorek, A.~Zeka, J.~Schwartz, B.~A. Coull, Measurement
  error caused by spatial misalignment in environmental epidemiology,
  Biostatistics 10~(2) (2009) 258--274.

\bibitem{leeshaddick}
D.~Lee, G.~Shaddick, Spatial modeling of air pollution in studies of its
  short-term health effects, Biometrics 66~(4) (2010) 1238--1246.

\bibitem{spde}
F.~Lindgren, H.~Rue, J.~Lindstr{\"o}m, An explicit link between gaussian fields
  and gaussian markov random fields: the stochastic partial differential
  equation approach, Journal of the Royal Statistical Society: Series B
  (Statistical Methodology) 73~(4) (2011) 423--498.

\bibitem{RueINLA}
H.~Rue, S.~Martino, N.~Chopin, Approximate bayesian inference for latent
  gaussian models by using integrated nested laplace approximations, Journal of
  the royal statistical society: Series b (statistical methodology) 71~(2)
  (2009) 319--392.

\bibitem{berrocal2010spatio}
V.~J. Berrocal, A.~E. Gelfand, D.~M. Holland, A spatio-temporal downscaler for
  output from numerical models, Journal of agricultural, biological, and
  environmental statistics 15~(2) (2010) 176--197.

\bibitem{liu2011empirical}
Z.~Liu, N.~D. Le, J.~V. Zidek, An empirical assessment of bayesian melding for
  mapping ozone pollution, Environmetrics 22~(3) (2011) 340--353.

\bibitem{wikle2005combining}
C.~K. Wikle, L.~M. Berliner, Combining information across spatial scales,
  Technometrics 47~(1) (2005) 80--91.

\bibitem{mcmillan2010combining}
N.~J. McMillan, D.~M. Holland, M.~Morara, J.~Feng, Combining numerical model
  output and particulate data using bayesian space--time modeling,
  Environmetrics: The official journal of the International Environmetrics
  Society 21~(1) (2010) 48--65.

\bibitem{sahu2010fusing}
S.~K. Sahu, A.~E. Gelfand, D.~M. Holland, Fusing point and areal level
  space--time data with application to wet deposition, Journal of the Royal
  Statistical Society: Series C (Applied Statistics) 59~(1) (2010) 77--103.

\bibitem{ciarlet2002finite}
P.~G. Ciarlet, The finite element method for elliptic problems, SIAM, 2002.

\bibitem{rinlabook}
M.~Blangiardo, M.~Cameletti, Spatial and spatio-temporal Bayesian models with
  R-INLA, John Wiley \& Sons, 2015.

\bibitem{RueMart}
H.~Rue, S.~Martino, Approximate bayesian inference for hierarchical gaussian
  markov random field models, Journal of statistical planning and inference
  137~(10) (2007) 3177--3192.

\bibitem{cameletti2013spatio}
M.~Cameletti, F.~Lindgren, D.~Simpson, H.~Rue, Spatio-temporal modeling of
  particulate matter concentration through the spde approach, AStA Advances in
  Statistical Analysis 97~(2) (2013) 109--131.

\bibitem{Martetal}
T.~G. Martins, D.~Simpson, F.~Lindgren, H.~Rue, Bayesian computing with inla:
  new features, Computational Statistics \& Data Analysis 67 (2013) 68--83.

\bibitem{ruiz2012direct}
R.~Ruiz-C{\'a}rdenas, E.~T. Krainski, H.~Rue, Direct fitting of dynamic models
  using integrated nested laplace approximations—inla, Computational
  Statistics \& Data Analysis 56~(6) (2012) 1808--1828.

\bibitem{rueheld}
H.~Rue, L.~Held, Gaussian Markov random fields: theory and applications, CRC
  press, 2005.

\bibitem{liu2017incorporating}
Y.~Liu, G.~Shaddick, J.~V. Zidek, Incorporating high-dimensional exposure
  modelling into studies of air pollution and health, Statistics in Biosciences
  9~(2) (2017) 559--581.

\bibitem{bivand2015comparing}
R.~Bivand, G.~Piras, Comparing implementations of estimation methods for
  spatial econometrics, Journal of Statistical Software 63 (2015) 1--36.

\bibitem{lindgren2015bayesian}
F.~Lindgren, H.~Rue, Bayesian spatial modelling with r-inla, Journal of
  statistical software 63 (2015) 1--25.

\bibitem{fuglstad2019constructing}
G.-A. Fuglstad, D.~Simpson, F.~Lindgren, H.~Rue, Constructing priors that
  penalize the complexity of gaussian random fields, Journal of the American
  Statistical Association 114~(525) (2019) 445--452.

\end{thebibliography}

\clearpage 
\appendix

\section{Components of the stage 1 model posterior distribution}
\label{app:stage1posteriors}

The form of each component is as follows:

\begin{align}
	\pi(\bm{w}|\bm{\chi},\bm{\xi},\bm{\theta}) &= \prod_{t=1}^T \pi(\bm{w}_t|\bm{\chi},\bm{\xi},\bm{\theta}) \\
	&\propto \prod_{t=1}^T |\sigma^2_e \mathbb{I}_M|^{1/2}\exp\bigg\{-\dfrac{1}{2} (\bm{w}_t-\bm{x}^*_{t,M})^\intercal (\sigma^2_e \mathbb{I}_M))^{-1} (\bm{w}_t-\bm{x}^*_{t,M}) \bigg\} \\
	&\propto (\sigma^2_e)^{-\tfrac{MT}{2}} \exp\bigg\{-\dfrac{1}{2\sigma^2_e} \sum_{t=1}^T (\bm{w}_t-\bm{x}^*_{t,M})^\intercal (\bm{w}_t-\bm{x}^*_{t,M}) \bigg\} 
\end{align}
\begin{align}
	 \pi(\tilde{\bm{x}}|\bm{\chi},\bm{\xi},\bm{\theta})&= \prod_{t=1}^T \pi(\tilde{\bm{x}}_t|\bm{\chi},\bm{\xi},\bm{\theta}) \\
	&\propto \prod_{t=1}^T |\sigma^2_{\delta} \mathbb{I}_G|^{1/2}\exp\bigg\{-\dfrac{1}{2} (\tilde{\bm{x}}_t- \alpha_0\mathbf{1}_G - \bm{x}^{*}_{t,P} )^\intercal (\sigma^2_{\delta} \mathbb{I}_G))^{-1} (\tilde{\bm{x}}_t- \alpha_0\bm{1}_G - \bm{x}^{*}_{t,P}\bm) \bigg\} \\
	&\propto (\sigma^2_{\delta})^{-\tfrac{GT}{2}} \exp\bigg\{-\dfrac{1}{2\sigma^2_{\delta}} \sum_{t=1}^T (\tilde{\bm{x}}_t- \alpha_0\bm{1}_G - \alpha_1\bm{x}^{*}_{t,P} )^\intercal (\tilde{\bm{x}}_t- \alpha_0\bm{1}_G - \alpha_1\bm{x}^{*}_{t,P} ) \bigg\} 
\end{align}

For the pseudo-zeroes, we have 
\begin{align}
\mathbf{0}_t|\bm{\chi},\bm{\xi},\bm{\theta} \sim N(-\bm{x}_t + \beta_{0}\bm{1}_{M+G} + \beta_{1}\bm{z}_t+ \bm{\xi}_t, \tfrac{1}{\tau_0}\mathbb{I}_G),
\end{align} 
where $\tau_0$ is a precision parameter and is fixed at a large value because of the absence of measurement error in the pseudo-zeroes. Thus, we have
\begin{align}
	 \pi(\mathbf{0}|\bm{\chi},\bm{\xi},\bm{\theta})&= \prod_{t=1}^T \pi(\mathbf{0}_t|\bm{\chi}_t,\bm{\xi}_t,\bm{\theta}_t)\\
	&\propto \bigg(\dfrac{1}{\tau_0}\bigg)^{-\tfrac{GT}{2}} \exp\bigg\{-\dfrac{\tau_0}{2} \sum_{t=1}^T (\bm{x}_t - \beta_{0}\mathbf{1}_{M+G} - \beta_{1}\bm{z}_t- \bm{\xi}_t )^\intercal (\bm{x}_t - \beta_{0}\mathbf{1}_{M+G} - \beta_{1}\bm{z}_t- \bm{\xi}_t) \bigg\}.
\end{align}

The form of the distribution of $\bm{\xi_t}|\bm{\theta}$ uses the fact that $\bm{\xi}_t|\bm{\xi}_{t-1}\sim N(\varsigma \bm{\xi}_{t-1},\bm{\Sigma}), t=2,\ldots,T$, and that $\bm{\xi}_1 \sim N(\mathbf{0}, \tfrac{1}{1-\varsigma^2}\bm{\Sigma})$. This gives
\begin{align}
	\pi(\bm{\xi}|\bm{\theta}) &= \pi(\bm{\xi}_1|\bm{\theta}) \prod_{t=2}^{T} \pi(\bm{\xi}_t|\bm{\xi}_{t-1} ,\bm{\theta}) \\
	&\propto |\tfrac{1}{1-\varsigma^2}\bm{\Sigma}|^{-\tfrac{1}{2}} \exp\bigg\{ -\dfrac{1}{2} \bm{\xi}_1^\intercal (\tfrac{1}{1-\varsigma^2}\bm{\Sigma})^{-1} \bm{\xi}_1\bigg\}  \times \\
	&\;\;\; \prod_{t=2}^T |\bm{\Sigma}|^{-\tfrac{1}{2}} \exp \bigg\{ -\dfrac{1}{2} (\bm{\xi}_t-\varsigma \bm{\xi}_{t-1})^\intercal (\bm{\Sigma})^{-1} (\bm{\xi}_t-\varsigma \bm{\xi}_{t-1}) \bigg\}.
\end{align}

\section{Plots of bias in the health model parameters} \label{app:plot.bias.poisson}

\begin{figure}[ht]
        \centering
     	\includegraphics[trim={3cm 3cm 3cm 3cm},clip,width=.7\textwidth]{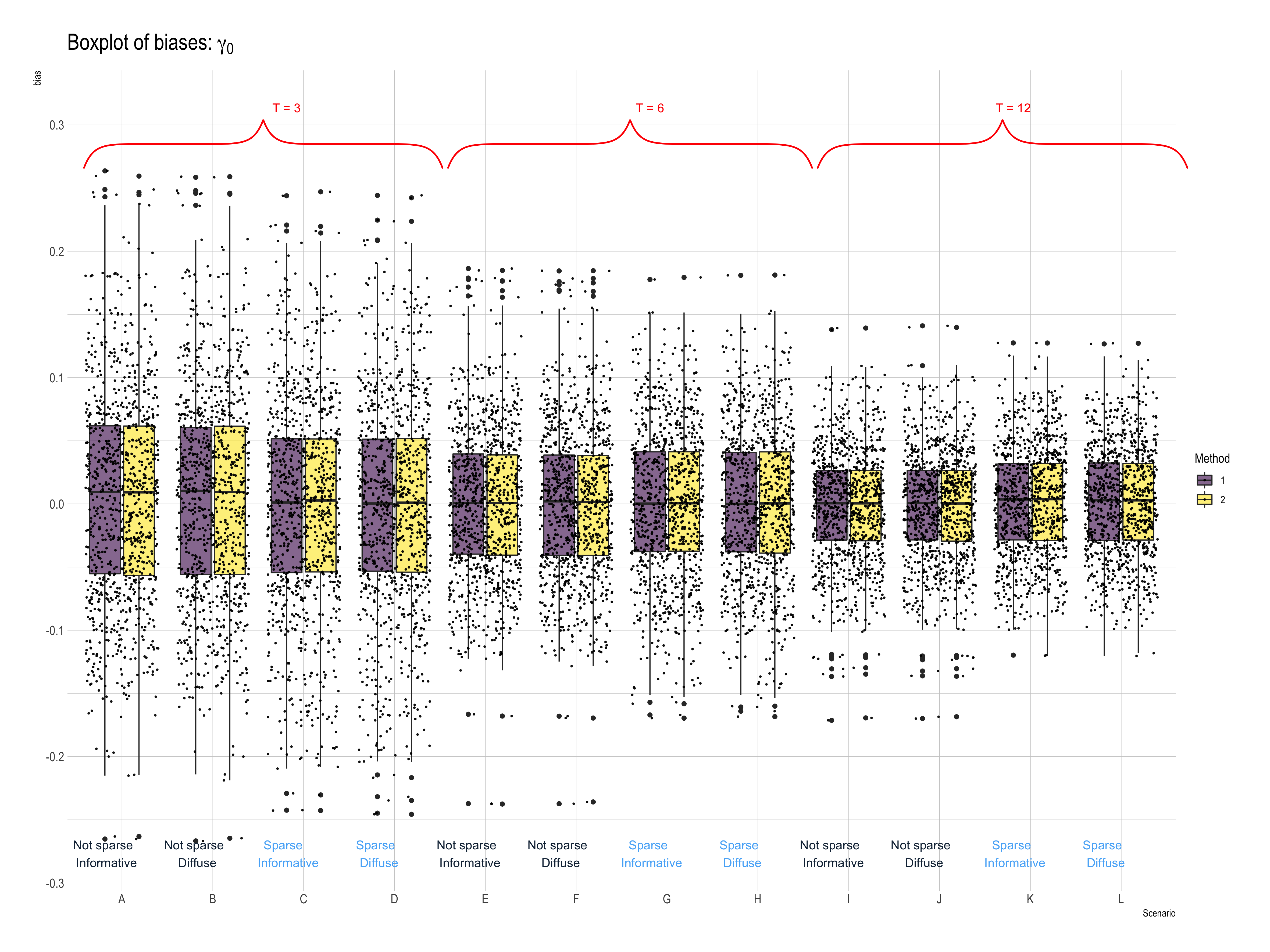}
	\caption{Plot of bias in $\gamma_0$ for all scenarios}
	\label{fig:Bias.perf.gamma0}
\end{figure}

\begin{figure}[ht]
        \centering
     	\includegraphics[trim={3cm 3cm 3cm 3cm},clip,width=.7\textwidth]{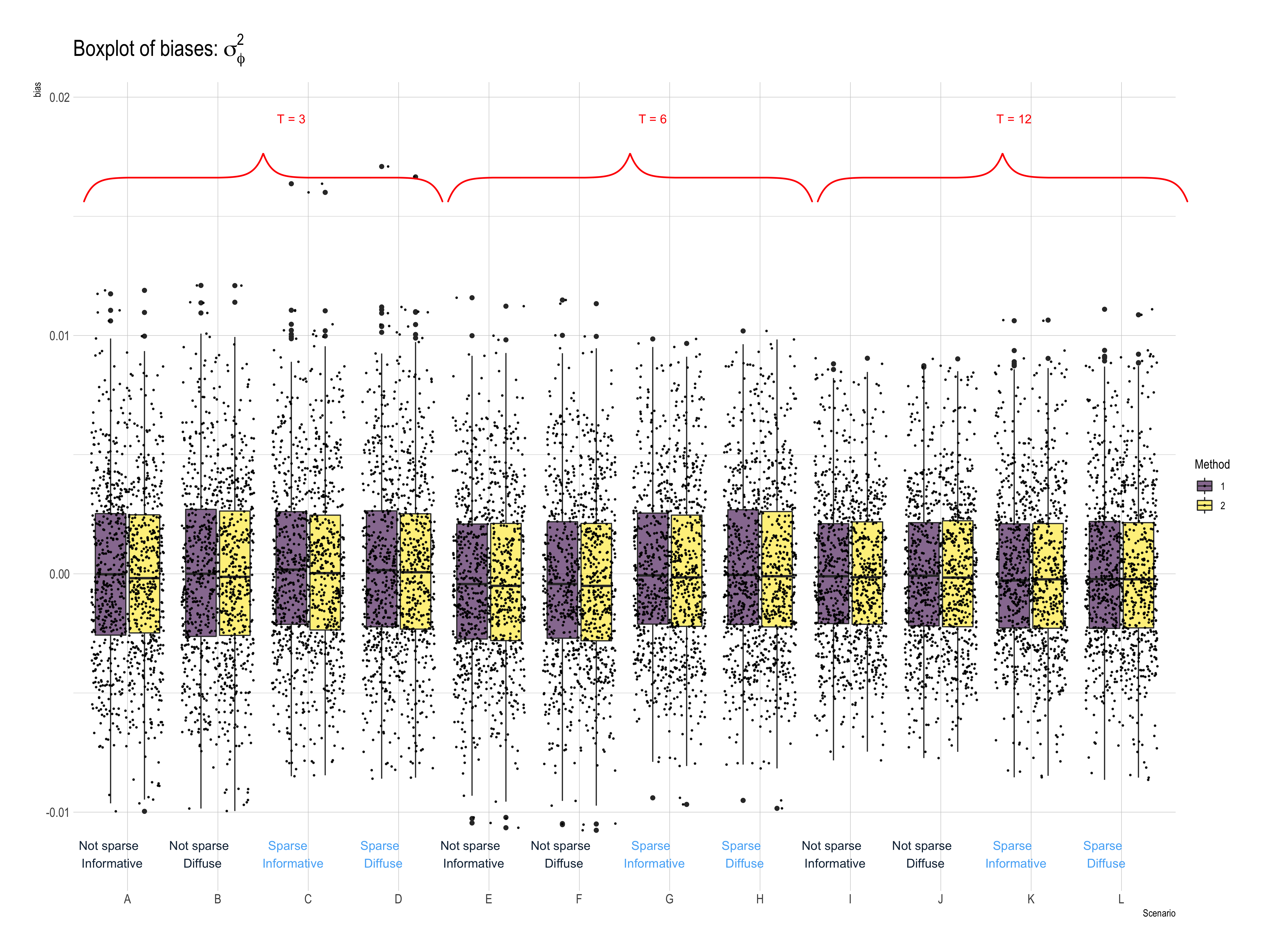}
	\caption{Plot of bias in $\sigma^2_{\phi}$ for all scenarios}
	\label{fig:Bias.perf.sigma2iid}
\end{figure}

\begin{figure}[ht]
        \centering
     	\includegraphics[trim={3cm 3cm 3cm 3cm},clip,width=.7\textwidth]{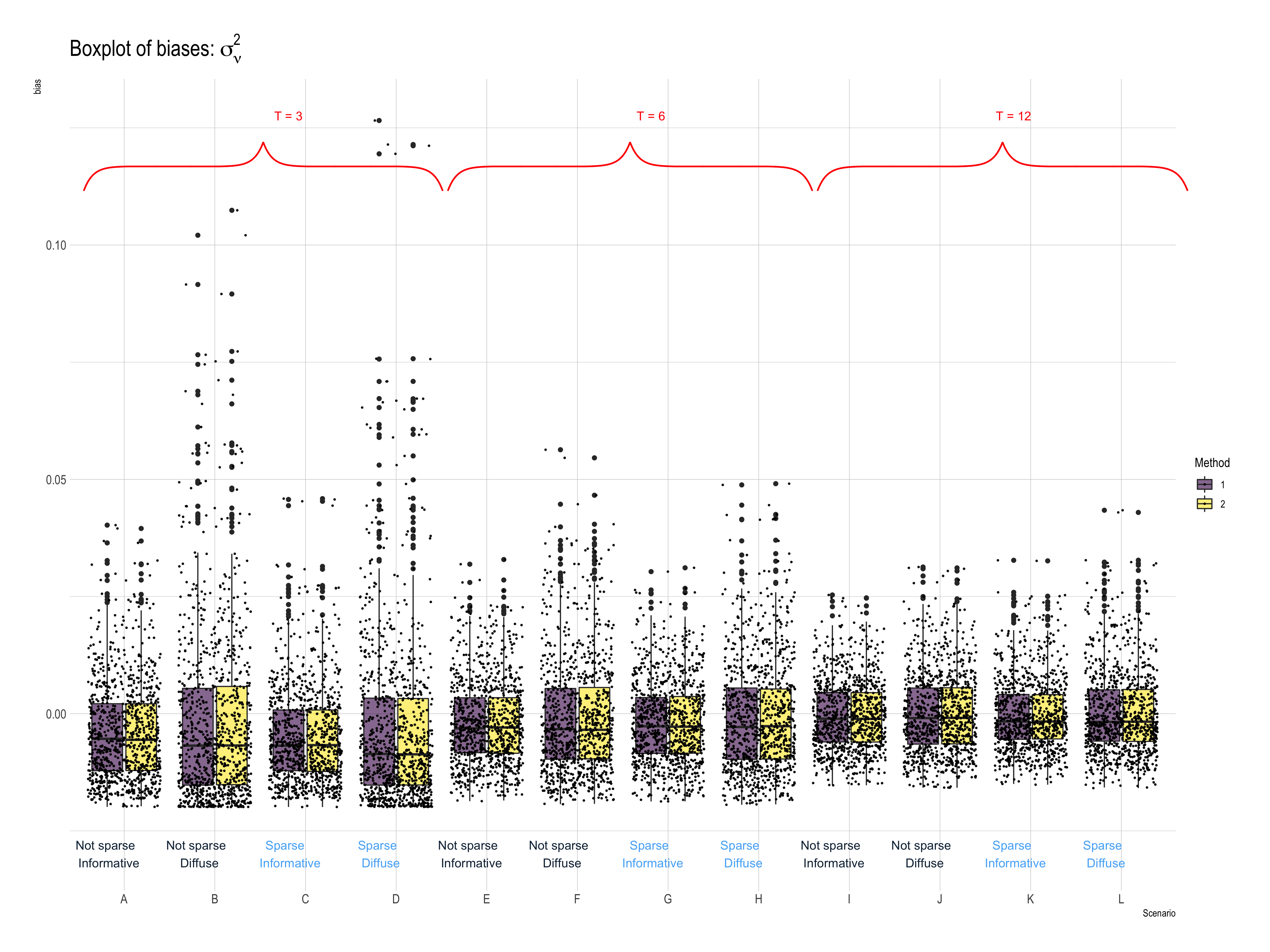}
	\caption{Plots of bias in $\sigma^2_{\nu}$ for all scenarios}
	\label{fig:Bias.perf.sigma2time}
\end{figure}

\clearpage

\section{Plot of RMSE in the health model parameters} \label{app:plot.rmse.poisson}

\begin{figure}[ht]
        \centering
     	\includegraphics[trim={3cm 3cm 3cm 3cm},clip,width=.7\textwidth]{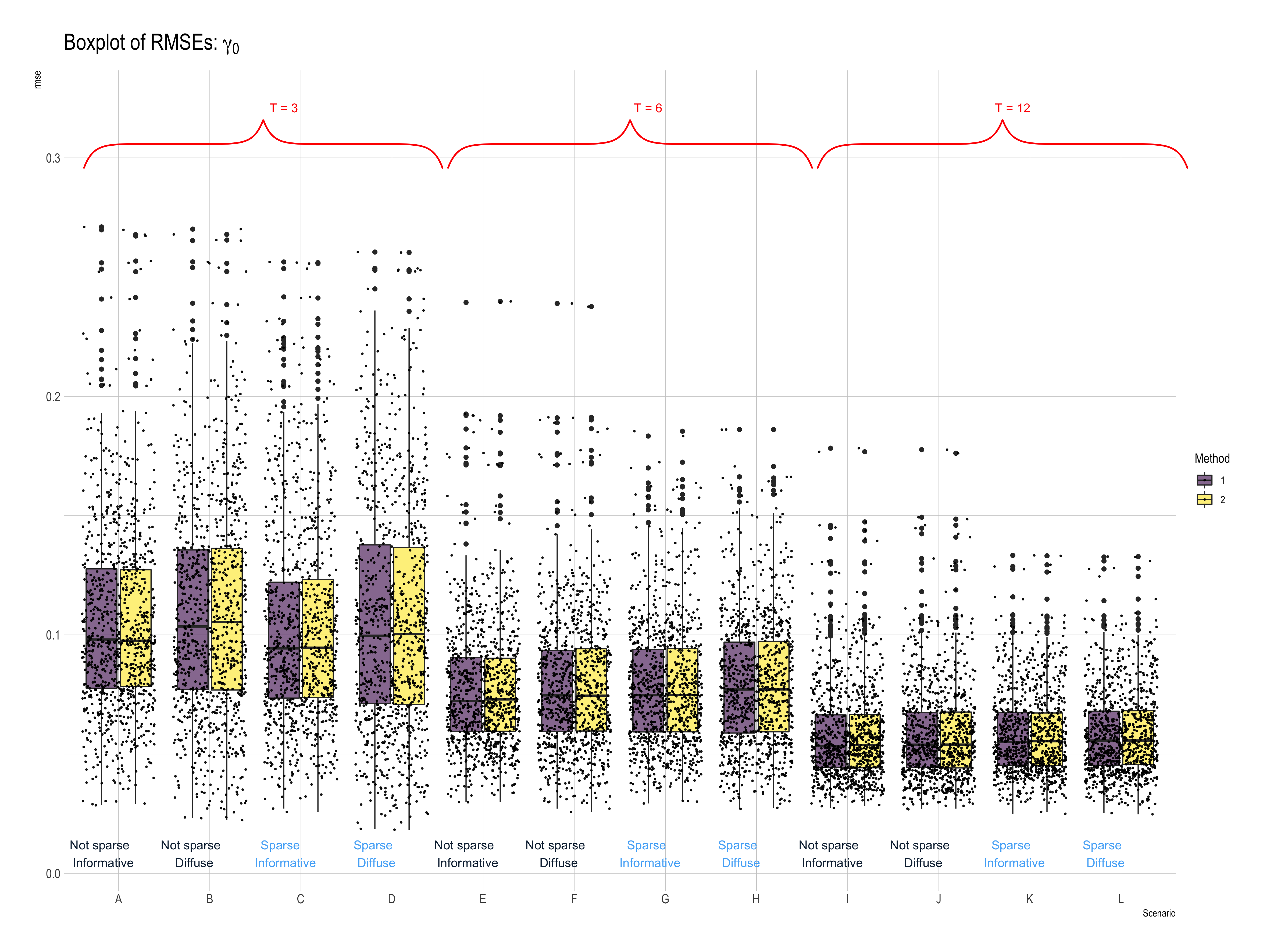}
	\caption{Plot of RMSE in $\gamma_0$ for all scenarios}
	\label{fig:rmse.perf.gamma0}
\end{figure}

\begin{figure}[ht]
        \centering
     	\includegraphics[trim={3cm 3cm 3cm 3cm},clip,width=.7\textwidth]{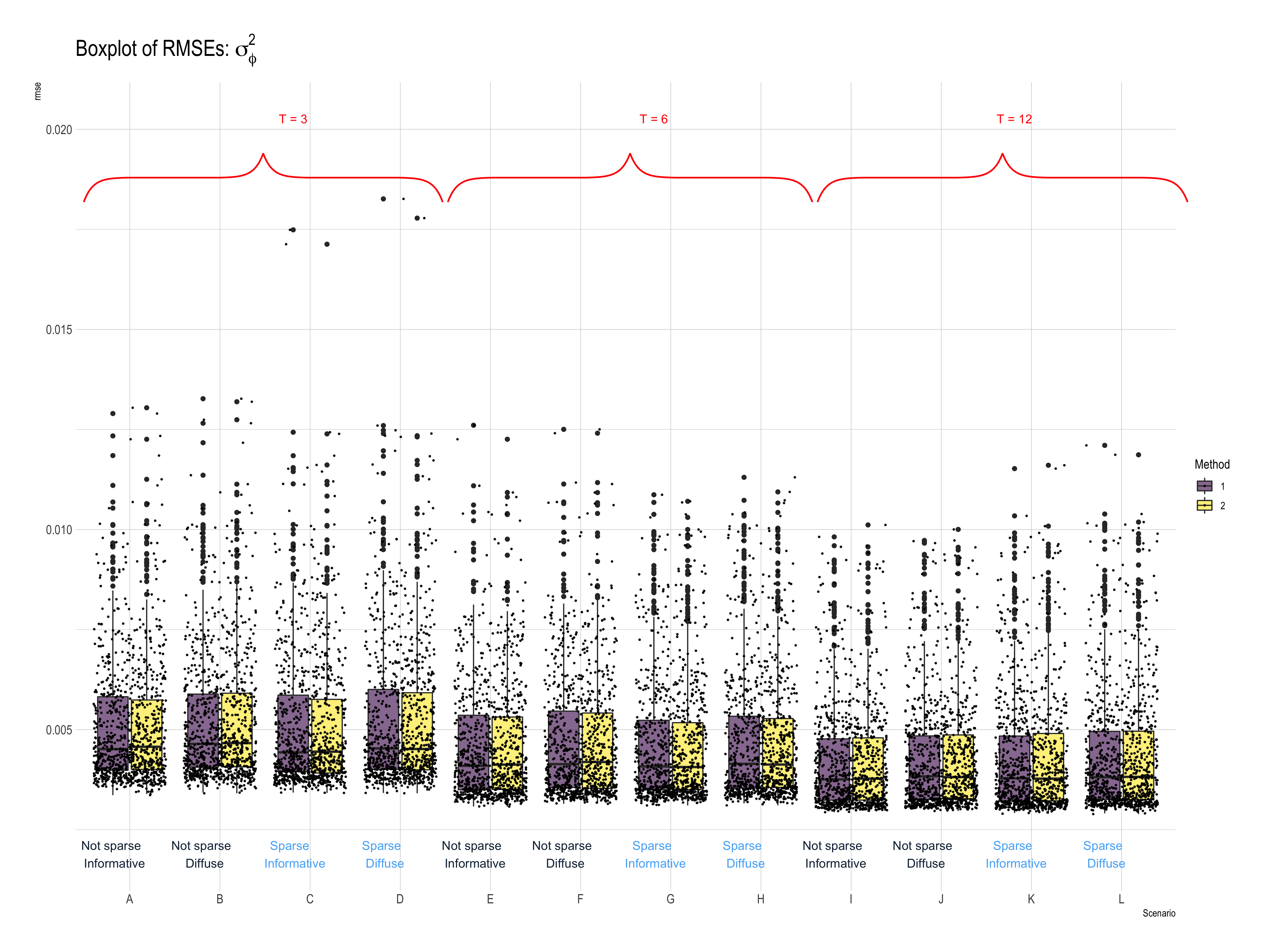}
	\caption{Plot of RMSE in $\sigma^2_{\phi}$ for all scenarios}
	\label{fig:rmse.perf.sigma2iid}
\end{figure}

\begin{figure}[ht]
        \centering
     	\includegraphics[trim={3cm 3cm 3cm 3cm},clip,width=.7\textwidth]{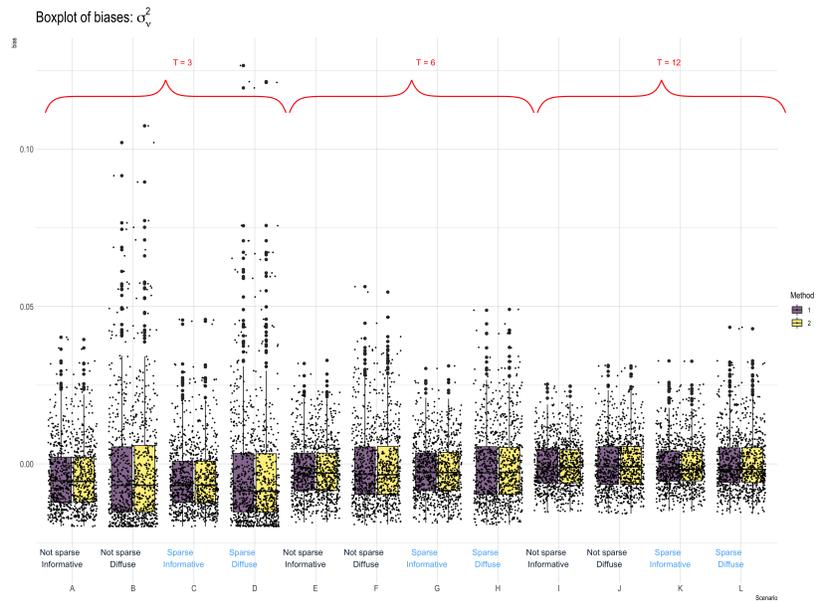}
	\caption{Plots of RMSE in $\sigma^2_{\nu}$ for all scenarios}
	\label{fig:rmse.perf.sigma2time}
\end{figure}


\end{document}